# Chandra X-ray Observations of NGC 1316 (Fornax A)

## Dong-Woo Kim
## G. Fabbiano

**Smithsonian Astrophysical Observatory**

(November 29, 2002)

Abstract

We report the results of the *Chandra* ACIS sub-arcsecond resolution X-ray observation of the archetypal merger radio galaxy NGC 1316 (Fornax A). We confirm the presence of fine sub-structures in the hot Interstellar Medium (ISM). Some of these are likely to result from interaction with the radio jets, while others may be related to a complex intermingling of different phases of the ISM. We detect a low-luminosity X-ray AGN with Lx = 5 x $10^{39}$ erg $sec^{-1}$ (in 0.3-8 keV) and a $\Gamma$=1.7 power-law energy spectrum. We also detect 81 point sources within the $25^{th}$ magnitude isophotal ellipse of NGC 1316 (Lx in the range of 2 x $10^{37}$ – 2 x $10^{39}$ erg $sec^{-1}$), with hard (kT~5 keV) X-ray spectra, typical of X-ray binaries, and a spatial radial distribution consistent with that of the optical (i.e., stellar) surface brightness. We derive the X-ray luminosity function (XLF) of these sources, correcting for the incompleteness at the faint end caused by the presence of the diffuse emission from the hot ISM in the central regions of NGC 1316 and by the widening of the *Chandra* PSFs at increasing distance from the aim point. With these corrections, the XLF is well reproduced by a single -unbroken- power law with a slope of -1.3 down to our threshold luminosity of ~3 x $10^{37}$ erg $sec^{-1}$. The hot ISM has temperatures in the 0.5 – 0.6 keV range, its surface brightness distribution is more centrally concentrated than that of the point sources, and its temperature appears to decrease at larger radii. These properties suggest that the ISM may be subject to partial winds. Taking into account the spectral complexity of the ISM, and the presence of unresolved low luminosity X-ray sources (which can be inferred from the spectra), we constrain the metal abundance of the hot ISM to be Z = 0.25 – 1.3 solar (90% confidence).

# 1. Introduction

The radio galaxy NGC 1316 (Fornax A) is a disturbed elliptical galaxy with numerous tidal tails. To explain this morphology, Schweizer (1980) suggested several low-mass, gas-rich mergers occurring over the last 2 Gyr (see also Ekers et al. 1983; Kim, Fabbiano and Mackie 1998, hereafter KFM; Mackie & Fabbiano 1998). NGC 1316 has been extensively observed in a wide range of wavelengths (see a summary of previous observations in KFM). In the radio band, Fornax A is the 3$^{rd}$ brightest object in the sky, with giant radio lobes (Wade 1961; Ekers et al. 1983), separated by ~200 kpc, consisting of polarized filaments (Fomalont et al. 1989), and S-shaped nuclear radio jets (Geldzahler and Fomalont 1984). In X-rays, it was observed with *Einstein* (Fabbiano et al. 1992), ROSAT PSPC (Feigelson et al. 1995), ASCA (Kaneda et al. 1995; Iyomoto et al. 1998) and ROSAT HRI (KFM). In particular, the ROSAT observations revealed ~10$^9$ M⊙ of hot ISM, with an inhomogeneous distribution. The sub-structures of the ISM were not highly significant, given the ROSAT HRI data quality, but nevertheless suggested a multi-phase ISM interacting with the radio jets (KFM). NGC 1316 is the first elliptical galaxy for which this type of jet/hot ISM interaction has been reported. Previous evidence of this phenomenon had been limited to galaxy clusters (Cygnus A, Carilli, Perley and Harris 1994; the Perseus cluster, Bohringer et al. 1993). *Chandra* observations are now revealing many other examples in several other elliptical galaxies and clusters (e.g., NGC 4374, Finoguenov and Jones 2001; and Hydra A, McNamara et al. 2000).

In this paper we report the results of a high spatial resolution, spectrally resolved *Chandra* ACIS (Weisskopf et al. 2000) observations of NGC 1316. The motivation of this work is two-fold. First, we attempt to confirm and study in detail the features of the hot ISM suggested by the ROSAT HRI (KFM). Second, NGC 1316 is an X-ray-faint elliptical (Fabbiano et al. 1992), where we would expect a large contribution to the X-ray emission to originate from low mass X-ray binaries (LMXBs, eg., Trinchieri & Fabbiano 1985; Canizares, Fabbiano & Trinchieri 1987; Kim, Fabbiano & Trinchieri 1992; Eskridge, Fabbiano & Kim 1995a, b; Fabbiano, Kim & Trinchieri 1994a; Pellegrini & Fabbiano 1994). With *Chandra* we can detect individual LMXBs (e.g., Sarazin et al. 2000), study their properties, and quantitatively estimate their contribution to the diffuse emission. We can also significantly reduce the bias introduced by the presence of unresolved sources in past discussions of the spectral and spatial characteristics of the ISM.

This paper is organized as follows: In section 2, we describe the *Chandra* observations, the data reduction and the results of spatial and spectral analyses. In section 3, we derive the XLF of point sources associated with NGC 1316 and estimate the uncertainties deriving from the varying detection thresholds across the field. In section 4, we summarize various emission components. In section 5, we discuss the implications of our results on the nature and properties of the hot ISM and its interaction with radio jets and other phases of ISM. Finally we summarize our conclusions in section 6.

We adopt a distance D = 18.6 Mpc throughout this paper, based on the HST measurements of Cepheid variables (Madore et al. 1999) in NGC 1365. At the adopted distance, 1" corresponds to 90 pc.

## 2. *Chandra* Observations and Data Analysis

NGC 1316 was observed for 30 ksec on April 17, 2001 (OBSID = 2022), with the *Chandra* Advanced CCD Imaging Spectrometer (ACIS; Garmire 1997). We used the back-illuminated CCD S3 (ccdid=7) because of its sensitivity at the low energies. To include NGC 1317 (6.3 arcmin away from NGC 1316) in the same S3 chip, a small offset was applied to the SIM (Science Instrument Module) position. NGC 1316 was kept close to on-axis to achieve the best spatial resolution.

The data were reduced with XPIPE (Kim et al. 2002), a custom pipeline which was specifically developed for the *Chandra* Multi-wavelength Project (ChaMP). XPIPE takes the CXC pipeline Level 2 data products and then applies additional data corrections (eg., gain correction, removing bad pixels/columns) and additional data screening (eg., removing background flares), performs source detection and determines source properties, such as flux, X-ray colors, variability and extent. Specifically for this observation, we have identified 2 bad pixels in CCD S2 (or ccdid=6), which resulted in 5 spurious sources. The CXC pipeline processing was done before the new S3 gain file was released on Sep. 2001. We have rerun **acis_process_events** to correct the ACIS gain and used follow-up tools available in CIAO (http://asc.harvard.edu/ciao/) to generate CXC pipeline-like products. Background flares are often seen in back-illuminated (BI) chips (CCD S1 and S3). They are excluded when the background rate is beyond $3\sigma$ from the average rate, which is measured iteratively after initial exclusion of flares. Removal of background flares reduced the effective exposure time of CCD S3 to 24.7 ksec.

To detect X-ray sources, **wavdetect** (Freeman et al. 2002), a wavelet detection algorithm available in CIAO was used. **wavdetect** is more reliable in finding individual sources (and not detecting false sources) in a crowded field containing extended emission than the traditional sliding box **celldetect** algorithm, although it requires a longer processing time. We set the **wavdetect** significance threshold parameter to be $10^{-6}$, which corresponds to 1 possibly spurious source and the scale parameter to cover 7 steps between 1 and 64 pixels. This made us sensitive to sources ranging from point-like to 32" in size, and in particular accommodates the variation of the PSF as a function of the off-axis angle. **wavdetect** tends to detect spurious sources near the detector edge. To avoid such false detections, we have utilized an exposure map (made at 1.5 keV) and applied a 10% exposure threshold. The exposure map was tailored to each CCD, based on the aspect history.

To extract source properties (such as count rates, spectra, etc.), we have used the 95% encircled energy (at 1.5 keV) radius centered at the **wavdetect** centroid, with a minimum of 3 arcsec to accommodate the radial variation of PSF. Background counts were determined locally for each source from an annulus from 2 to 5 times the source radius, after excluding nearby sources.

## 2.1 Point Sources

The detected X-ray sources are shown in Figure 1 superimposed on the X-ray (left panel) and on the Digitized Sky Survey (http://stdatu.stsci.edu/dss) optical (right panel) images. In this figure (and throughout this paper), north is to the top and east is to the left of the image. Extended sources are found at the locations of NGC 1316 and NGC 1317 as seen in ROSAT data (KFM). In addition, the *Chandra* observations reveal 94 sources, 83 of them in CCD S3. 81 sources (77 in S3 and 4 in S2) are within the $D_{25}$ ellipse (taken from RC3) of NGC 1316. The source density increases toward the center of NGC 1316, indicating that most of them are related to NGC 1316. Three sources are found within $D_{25}$ of NGC 1317 with the brightest, extended one at the center of NGC 1317. The list of sources (including sources found in other CCDs) is given in Table 1. The net source counts estimated by XPIPE (ie., by aperture photometry as described above) are usually in good agreement with those determined by **wavdetect**. Exceptions are those sources within 20 arcsec from the center of NGC 1316 where the extended diffuse X-ray emission makes it hard to accurately measure source and background counts. Consequently, the net counts and rates are subject to a large uncertainty and these sources are marked in the Table (they are not used below in determining XLF and in spectral fitting).

Also marked in the table are those sources which have one or more sources within their source extraction radius, introducing additional uncertainty on their count rates. Three sources are found at the edge of the chip and their counts should be considered as a lower limit. These are marked in Table 1. After correcting for effective exposure and vignetting, the X-ray flux in the 0.3-8.0 keV band is calculated with an energy conversion factor, ECF, assuming a power-law source spectrum with $\Gamma_{ph} = 1.7$ and $N_H = 3 \times 10^{20}$ cm$^{-2}$: ECF= $6.037 \times 10^{-12}$ erg cm$^{-2}$ sec$^{-1}$ per 1 count sec$^{-1}$ for the BI chips and $9.767 \times 10^{-12}$ erg cm$^{-2}$ sec$^{-1}$ for the front-illuminated (FI) CCD chips. With the adopted distance of 18.6 Mpc, the X-ray luminosities of the point sources range from a ~$2 \times 10^{37}$ to a ~$8 \times 10^{39}$ erg sec$^{-1}$. We have identified one possible, off-center, Ultra Luminous X-ray source (ULX) with $L_x = 2.4 \times 10^{39}$ erg sec$^{-1}$ (CXOU J032251.2-370949). Following Zezas and Fabbiano (2002), we define the ULX as a non-AGN (i.e., off-center) point source with $L_x > 2 \times 10^{39}$ erg sec$^{-1}$. This source is marked in Figure 1 and also in Table 1. No optical counterpart is seen in the DSS image (see Figure 1). There are 3 more sources with net-count > 200 (or, $L_x$ > a few x $10^{39}$ erg sec$^{-1}$) within the $D_{25}$ ellipse of NGC 1316 (see Table 1). Two of them are corresponding to the elongated feature at the center of NGC 1316 (see section 2.2). The 3$^{rd}$ source, CXOU J032241.2-371235, is 10" SW from the center, hence its X-ray flux is highly uncertain because of the strong diffuse X-ray emission. Counting pixel values manually suggests that its count might be over-estimated by as much as ~50%. Five sources are found to be coincident with globular clusters identified in HST images (Shaya, 1996) and in ground observations (Goudfrooij et al. 2001). They are marked as such in Table 1. Also identified is one potential super soft X-ray source (SSS), CXOU J032235.9-371135, for which all X-ray photons are in the soft band (0.3-0.9 keV).

Based on the Log N – Log S of *Chandra* Deep Field sources (Brandt et al. 2001; Giacconi et al. 2001) and ChaMP sources (Kim et al, 2002), we expect less than 7 serendipitous sources within $D_{25}$ of NGC 1316 in S3. The radial distribution of point sources is also consistent with the radial profile of the optical light, i.e., the stellar distribution (see section 2.3), suggesting that most of them are LMXBs associated with NGC 1316. The spectral properties of these sources are also consistent with this hypothesis (section 2.4.1).

## 2.2 Surface Brightness Distribution of the Diffuse X-ray Emission

To identify the substructures of extended, diffuse X-ray emission, we have applied **csmooth**, an adaptive smoothing algorithm available in CIAO. Figure 2 shows the smoothed image within the central 3 x 3 arcmin region. In this arcmin scale image, the diffuse X-ray emission is elongated in the north-south direction, and does not follow either the major axis (PA=50°) or the minor axis of NGC 1316. The ellipse in the center of Figure 2 indicates the optical figure of NGC 1316 with its size equal to 1/10 of $D_{25}$. Also noticeable is a secondary maximum of the surface brightness ('blob') at 30-60" north of the center. This feature is extended when compared to the PSF and its X-ray emission is possibly soft (see section 2.4.3). We address the nature of these large scale N-S elongations in section 5.

In the sub-arcmin scale, the X-ray surface brightness is elongated along the major axis of the optical figure, but it is far from featureless. The enlarged view of the central region of NGC 1316 is shown in Figure 3. It is clearly noticeable that the X-ray surface brightness distribution forms valleys (cavities) toward SE (PA = 120°) and NW (PA = 315°). These are the directions along which the radio jets propagate (see Figure 2 in Geldzahler and Fomalont 1984). In Figure 3, red lines indicate the directions and approximate sizes of the radio jets (see section 5.2 for further discussions on the interaction between the hot ISM and radio jets). It was one of the main purposes of this observation to confirm the reality of these X-ray valleys, that were suggested by the ROSAT HRI image (KFM). If we compare the counts within a 3" radius circle in the X-ray valley and in the surrounding region at a similar distance from the center, the difference is significant at a ~10 σ level. The central 3" region exhibits a double-peak feature, elongated toward NE (see also Figure 4).

Figure 4 shows the central 3' x 3' smoothed X-ray images in different energy bands: soft (0.3-0.9 keV), medium (0.9-2.5 keV), and hard (2.5-8.0 keV). The diffuse emission is most prominent in the soft band as expected from the hot gaseous X-ray emission. Also the northern blob is most clearly seen in the soft band (see below for more discussion). The point sources instead are seen more distinctly in the medium band as expected for LMXBs with a harder spectrum. A point source at the nucleus is unambiguously identified in the hard band. The location of the nucleus determined from this hard band image is marked as a circle at the center of Figure 3. A true color image made by combining the three band images (red for the soft band, green for the medium band and blue for the hard band) is shown in Figure 5.

## 2.3 Radial Profiles of Diffuse X-ray Emission and Point Sources

We have derived radial surface brightness profiles to study the radial distribution of the hot gas and point-source components and compare them with the optical (stellar) distribution. Given the obvious lack of circular symmetry of the diffuse X-ray emission, we have extracted radial profiles from different angular sectors, after excluding detected point sources. Beta models, $\Sigma_x \sim (1 + (r/a)^2)^{-3\beta+0.5}$, were then used to model these profiles. Toward the E-W direction at PA = 30° – 150° and 210° – 330°, excluding the northern blob and the less prominent southern extension, the radial profile is smooth and the best-fit β is 0.57 – 0.59 at 90% confidence, corresponding to a radial slope of -2.4 – -2.5. Toward the N-S direction, however, the radial profile steepens around 20-30" and then flattens because of the elongated features. The best-fit β is 0.43 – 0.49 at 90% confidence which corresponds to a radial slope of -1.6 – -1.9. The X-ray radial profile measured toward the E-W direction is shown in Figure 6 where observed data points are marked by open circles with error bars and the best-fit beta model is represented by the dashed line. Although the data are well fitted in most radial bins, there is a significant deviation from the beta model at the nucleus, which requires an additional point source. This nuclear source is most prominent in the hard band X-ray image (Figure 4). We have used a CIAO tool, **mkpsf** to reproduce a point source at the center. The intensity of this nuclear source is consistent with the results of the X-ray spectral fitting (see section 2.4.2). The solid line in Figure 6 indicates the composite beta model + central point source profile. The central source profile is marked by the dotted line in the same figure. The estimated X-ray flux of the nuclear component is $1.2 \times 10^{-13}$ erg sec$^{-1}$ (with a 15% error) in the broad energy band (0.3-8 keV).

Also plotted in the Figure 6 (in green) is the radial profile of the stellar light distribution from the HST I band (Shaya et al. 1996). The I band profile decreases radially as $r^{-1.16}$ and is much flatter than the radial distribution of the diffuse X-ray emission. The X-ray point sources detected in the *Chandra* observations (red squares in Figure 6) are distributed like the stellar light, as it would be expected from LMXBs in NGC 1316, instead of following the diffuse X-ray emission. This means the contribution from LMXBs to the X-ray emission becomes more significant at the outer radii. It has an important implication when considering the effect of hidden populations of LMXBs in X-ray faint early-type galaxies observed with lower angular resolution than *Chandra*. In particular, lower resolution data would give a biased distribution of the hot ISM, and wrong (in excess) mass measurements, if the contribution of LMXBs is ignored. Ignoring the LMXBs would also affect spectral results (see Section 5 for further discussion).

## 2.4 Spectral Analysis

Following the science threads in http://cxc.harvard.edu/ciao/documents_threads.html, we have produced response files (rmf and arf) to be used in the spectral fitting. We have also applied a new method to produce weighted response files to take into account the variation with detector location (as described in the science thread available in http://asc.harvard.edu/ciao/), but the results are almost the same. After this paper was submitted, a time-dependent degradation of the low energy ACIS Quantum Efficiency was reported (http://cxc.harvard.edu/cal/Links/Acis/acis/Cal_prods/qeDeg/index.html). We have regenerated the spectral response files and rerun all the spectral fitting. We found that the spectral parameter most affected is $N_H$, which can become lower by as much as a factor of 2. Also the kT of the softer spectral component in multi-components fits becomes slightly lower. However, the main results remain the same. We report here the results from the latest analysis, performed with the best available calibration.

The background spectrum was extracted from five circular, source-free regions with a 60" radius in CCD S3, situated between 2' and 5' from the center of NGC 1316. Different choices of background regions within the same CCD do not change our results, because point sources (the nucleus and LMXBs) are well within very small source regions and the diffuse emission is extended only to ~1-2'. We have fit, within an energy range of 0.3-5.0 keV, emission models including a thermal gas model (MEKAL), Bremsstrahlung and power-law, alone and in combination. For solar abundances, we take the photospheric values given in Anders and Grevesse (1989), which is about 40% larger than the meteoric value in Fe. The spectral fitting results are summarized in Table 2-4. We discuss these results below, for the different galaxian components.

**2.4.1 Point Sources**

Since most point sources are not bright enough for individual spectral analysis, we have co-added their spectra to determine their average spectral properties. We have used point sources only at galactocentric radii r > 20" to avoid confusion with the strong diffuse emission near the center of NGC 1316. We have only used sources found in CCD S3, to avoid the complexity introduced by different detector responses. We find that the observed spectra of point sources are well fit with single-component models, either a thermal model with best-fit temperature of kT~4-6 keV or with a power-law model with the best-fit photon index of $\Gamma_{ph}$ ~2.0 (Table 2). This power-law spectrum is softer than those determined in a few other early type galaxies (eg. NGC 4697, Sarazin et al. 2001: $\Gamma_{ph}$ ~1.6). However, if $N_H$ is fixed at the galactic value, a softer $\Gamma_{ph}$ ~1.8 is obtained, more in keeping – within the errors – with these other studies. The spread of X-ray colors of point sources is well within the range of those found in other galaxies (see below).

More conservatively, we have also used sources only at r > 60" where the diffuse emission reaches the background level. The results are almost the same. Excluding the brightest source (the ULX candidate, CXOU J032251.2-370949), to confirm that the average properties are not dominated by a small number of peculiar sources, does not affect the results.

We have also investigated the spectral properties of the sources by means of X-ray colors. In Figure 7, we plot individual point sources detected with more than 20 counts in a X-ray color-color diagram. Following the ChaMP convention, (also similar to those previously used in studying Einstein data by Kim et al. 1992), we have defined two X-ray colors to represent two spectral parameters, absorption and slope:

$$C21 = -\log(\text{counts in 0.9-2.5 keV}) + \log(\text{counts in 0.3-0.9 keV})$$
$$C32 = -\log(\text{counts in 2.5-8.0 keV}) + \log(\text{counts in 0.9-2.5 keV})$$

There are several reasons for this particular selection of energy bands (see Kim et al. 2002 for details). In particular, we want (1) isolate the two spectral parameters (i.e., C21 is mostly sensitive to the absorption and C32 to the spectral slope) and (2) distribute the spectral counts almost equally in the three energy bands for a typical source with $\Gamma_{ph} \sim 2.0$ so that we can achieve the highest possible statistical significance. In Figure 7, filled circles are for source with more than 50 counts, open circles for sources with counts between 30 and 50, and stars (without corresponding error bars) for sources with counts between 20 and 30. The grid indicates the predicted locations of sources with various photon indices (from 0 to 4) and absorption column densities (from $10^{20}$ to $10^{22}$ cm$^{-2}$). The points are scattered in the X-ray color-color diagram to the extents comparable to errors in X-ray colors with a centroid of their distribution around $\Gamma_{ph} \sim 2.0$ and $N_H \sim$ a few x $10^{20}$ cm$^{-2}$. This is consistent with the average properties obtained by fitting the combined spectra.

Table 2. Spectral Fitting of the Combined Point Source Spectrum

| model | $N_H^+$ | $KT/\Gamma_{ph}$ | Z | $\chi^2_{reduced}$ ($\chi^2$/dof) |
|---|---|---|---|---|
| (total counts = 1730) | | | | |
| 1 | 0.000 (0.01) | 5.57 (1.5) | 1.00 | 0.97 (= 76.7 / 79 ) |
| 2 | 0.000 (0.01) | 3.29 (1.0) | | 0.74 (= 58.1 / 79 ) |
| 3 | 0.054 (0.02) | 2.03 (0.1) | | 0.67 (= 53.1 / 79 ) |
| 4 | 0.025 | 1.85 (0.1) | | 0.69 (= 55.2 / 80 ) |

Errors in parentheses are at the 90% confidence level for one interesting parameter.
+ $N_H$ is in unit of $10^{22}$ cm$^{-2}$
Model 1: one-component model: MEKAL, abundance fixed to solar
Model 2: one-component model: Bremsstrahlung
Model 3: one-component model: power-law
Model 4: one-component model: power-law with fixed $N_H$

The filled circle surrounded by a larger open circle in Figure 7 indicates the brightest source, (CXOU J032251.2-370949, the ULX). Its spectrum appears to be slightly absorbed ($N_H \sim$ a few x $10^{21}$ cm$^{-2}$) and steep ($\Gamma_{ph} \sim 2.5$-4.0). Both spectral parameters are within the range of the ULXs found in the Antennae galaxies (Zezas et al. 2002; see Figure 13 in their paper). If we assume $\Gamma_{ph} = 3.5$ and $N_H = 2$ x $10^{21}$ cm$^{-2}$, then its luminosity increases by a factor of 2, ie., $L_x = 4.7$ x $10^{39}$ erg sec$^{-1}$. We do not see temporal variability of this source either by binning the data or by applying a Bayesian-block analysis to the unbinned data (developed by J. Drake for the ChaMP, private communication; Kim et al. 2002).

A very soft emission component (0.2-0.3 keV) was reported in very X-ray faint early type galaxies (Kim et al, 1992; Fabbiano et al. 1994; Kim et al. 1996), and its nature (from hot ISM or from stellar sources) has been debated (e.g. Pellegrini & Fabbiano 1994). It was also suggested that LMXBs might be the cause of this very soft emission, based on ROSAT and ASCA spectra of bright point sources in M31 and NGC 4697 (Irwin and Sarazin 1998; Irwin et al. 2000). *Chandra* data exclude this latter hypothesis (the present results; see also Irwin et al. 2002).

**2.4.2 Nucleus**

As discussed in sections 2.2 and 2.3, a hard nuclear point source is clearly detected in the spatial analysis of the region where the diffuse emission is also most intense. We have fitted the spectrum from the central 5" of NGC 1316 with a power-law model for the AGN plus a MEKAL model for the hot ISM. The spectrum is well reproduced by this two-component model (reduced $\chi^2$ of 0.65 for 59 degrees of freedom). The best-fit kT of the thermal component is 0.62 ($\pm$ 0.02) keV and the best-fit photon index of the AGN component is 1.76 ($\pm$ 0.2). The metal abundance of the hot ISM was assumed solar (this is consistent with the constraints set in 2.4.4 on the metal abundance). The contribution from the power-law component is $F_x$ = 1.3 ($\pm$ 0.2) x $10^{-13}$ erg sec$^{-1}$ cm$^{-2}$ in 0.3 – 8.0 keV, which is in excellent agreement with the estimate from the radial profile analysis (in section 2.3).

For comparison, spectral fitting with single-component models either with fixed or varying metal abundance were also performed and the results are shown in Table 3 and Figure 8. A single-component model with best-fit Z = 0.14 solar also gives a formally acceptable fit, with a reduced $\chi^2$ of 1.1 for 60 degrees of freedom. However, it is obvious from Figure 8 that the model predictions are systematically lower than the observed data at E > 2 keV. In contrast, the two-component model suggested by the image (thermal gas + power-law) not only has a lower reduced $\chi^2$, but also reproduces the observed data well in the full energy range. Given the signal to noise ratio of the data, Z cannot be constrained in this case. This demonstrates that the spectral results depend critically on the adopted emission model (as pointed out in e.g., Kim et al. 1996). We will address this point further later in this paper.

**2.4.3 Diffuse X-ray Emission: Hot ISM and Undetected LMXBs**

After excluding detected point sources, we have extracted spectra from the diffuse X-ray emission in 3 different annuli, with galactocentric radii between 5" and 60" (see Table 3). Because of the resolution of *Chandra*, the nuclear AGN does not contribute at r > 5" (Figure 6), and the diffuse X-ray emission in these annuli is likely to be mostly from the hot ISM and undetected LMXBs. The observed spectra are well reproduced by a thermal gas (MEKAL) for the hot ISM + Bremsstrahlung (or power-law) model for undetected LMXBs. The best-fit kT of the soft component has similar values in each annulus for all the applied models. While the kT of the innermost annulus is 0.63 keV (with a 0.03 keV uncertainty), consistent within the errors with that of the second annulus, our results suggest a general decrease of the temperature with increasing radius. We find kT=0.50 keV at r = 30''- 60". For all the fits, the reduced $\chi^2$ is between 0.7 and 1.0.

Table 3. Spectral Fitting of Diffuse X-ray Emission

A. Central Bin (r < 5")

| model | $N_H^+$ | $KT_1$ | Z | $\Gamma_{ph}$ | Fx* total | Fx* 1st-comp | Fx* 2nd-comp | $\chi^2_{reduced}$ ($\chi^2$/dof) |
|---|---|---|---|---|---|---|---|---|
| (r < 5" ; total counts = 1613) | | | | | | | | |
| 1 | 0.00 (0.01) | 0.65 (0.03) | 1.00 | | 1.84 | | | 2.06(= 125.8/61) |
| 2 | 0.11 (0.01) | 0.61 (0.03) | 0.13 (0.01) | | 2.05 | | | 1.13(= 67.7/60) |
| 4 | 0.02 (0.01) | 0.62 (0.03) | 1.00 | 1.76 (0.2) | 2.71 | 1.43 | 1.28 | 0.64(= 38.0/59) |

B. Outskirts (r = 5" - 60")

| model | $N_H^+$ | $KT_1$ | Z | $kT_2/\Gamma_{ph}$ | Fx* total | Fx* 1st-comp | Fx* 2nd-comp | $\chi^2_{reduced}$ ($\chi^2$/dof) |
|---|---|---|---|---|---|---|---|---|
| (r = 5"-15" ; total counts = 1851) | | | | | | | | |
| 1 | 0.00 (0.01) | 0.65 (0.02) | 1.00 | | 2.21 | | | 1.37(= 79.5/58) |
| 2 | 0.05 (0.02) | 0.63 (0.04) | 0.21 (0.04) | | 2.40 | | | 0.71(= 40.7/57) |
| 3 | 0.00 (0.01) | 0.64 (0.03) | 1.00 | 1.38 (1.5) | 2.52 | 1.82 | 0.70 | 0.70(= 39.0/56) |
| 4 | 0.05 (0.01) | 0.63 (0.03) | 1.00 | 2.74 (0.6) | 2.55 | 1.75 | 0.81 | 0.64(= 36.0/56) |
| (r = 15"-30" ; total counts = 1168) | | | | | | | | |
| 1 | 0.00 (0.01) | 0.54 (0.03) | 1.00 | | 1.08 | | | 1.37(= 64.2/47) |
| 2 | 0.00 (0.01) | 0.59 (0.03) | 0.26 (0.02) | | 1.23 | | | 1.03(= 47.6/46) |
| 3 | 0.00 (0.01) | 0.60 (0.04) | 1.00 | 0.45 (**) | 1.23 | 0.93 | 0.29 | 1.05(= 47.0/45) |
| 4 | 0.00 (0.01) | 0.59 (0.04) | 1.00 | 2.98 (0.8) | 1.27 | 0.95 | 0.31 | 1.06(= 47.5/45) |
| (r = 30"-60" ; total counts = 2520) | | | | | | | | |
| 1 | 0.00 (0.01) | 0.42 (0.04) | 1.00 | | 1.85 | | | 1.07 (= 83.5/78) |
| 2 | 0.00 (0.01) | 0.50 (0.04) | 0.22 (0.03) | | 2.11 | | | 0.81 (= 62.5/77) |
| 3 | 0.00 (0.01) | 0.52 (0.04) | 1.00 | 0.32 (**) | 2.12 | 1.60 | 0.52 | 0.80 (= 60.5/76) |
| 4 | 0.00 (0.01) | 0.51 (0.04) | 1.00 | 3.60 (1.0) | 2.16 | 1.64 | 0.52 | 0.81 (= 61.7/76) |

Errors in parentheses are at the 90% confidence level for one interesting parameter.
+ $N_H$ is in unit of $10^{22}$ cm$^{-2}$
* Fx is in unit of $10^{-13}$ erg s$^{-1}$ cm$^{-2}$ in a 0.3 - 8.0 keV energy band
** not constrained

Model 1: one-component model: MEKAL with fixed Z
Model 2: one-component model: MEKAL with varying Z
Model 3: two-component model: MEKAL + power-law
Model 4: two-component model: MEKAL + Bremsstrahlung

As it was shown in the case of the nuclear spectrum, single thermal component models may give a formally acceptable result for very low abundance values. However, a systematic residual at the high energies points out the presence of a hard LMXB contribution (see Figure 9, for the r = 5-15" annulus). We see a similar trend in the outer radial bins, but the excess at the higher energies is getting weaker. At radial bins of r = 15"-30" and r = 30"-60", the formal statistics of spectral fitting do not favor a two-component model over a one-component, low-abundance model. However, based on our analysis of the two inner bins, we adopt the two-component model.

In the two-component model, the hard component contributes about 25% of the total emission. In the region (r = 20" - 60"), where both X-ray emission components from the hot ISM and LMXBs can be measured, the X-ray contribution from the detected point sources is about 30% of the total X-ray emission. As discussed in section 3, the total X-ray luminosity of the hidden population of LMXBs can be 40% - 100% of that of the detected LMXBs, depending on the lower limit of XLF. Therefore, the expected contribution from the hidden LMXBs is about 12% - 30%, which is in good agreement with the result estimated by the spectral fitting (25%). The relative contribution from the hard component to the total X-ray emission is more or less constant in the 3 radial bins, indicating that the amount of the hard component decreases with increasing radius, as expected if it would come from LMXBs. However, due to the uncertainties, it is statistically indeterminate whether the radial decrease is steeper than that of the diffuse X-ray emission and as steep as that of the detected point sources (as discussed in section 2.3).

**2.4.4 Metal abundance of the hot ISM**

The one-component model fit gives a very low value for the metal abundance of the hot ISM (Z = 0.21, with a formal 90% error of 0.04), but as we discuss in the previous section this model does not represent well the higher energy data, where excess emission can be seen (fig. 9), suggesting an undetected population of LMXBs. The metal abundance of the ISM is not well constrained with the two-component model, where this hard component is also fitted to the data, although in all cases solar values of the metal abundance (Z) are allowed by the fit.

To improve the constraints on Z, we have performed spectral fitting by (1) fixing the parameters of the harder spectral component, using the values determined in section 2.4.1 for the integrated point source spectrum; and (2) combining spectra from different annuli. In both cases, we obtain an acceptable range for Z of ~0.25 – ~1.5 solar. However, both these approaches disregard the spectral complexity that is suggested by our radial-annuli-based analysis. As shown in Table 3, although the temperature of the soft component does not change much ($\Delta kT \sim 0.12$ keV in 3 radial bins between 5" and 60"), the temperature change is statistically significant at the 6σ level, given the very small error of the measured temperature in each bin ($\varepsilon_{kT} = 0.03 – 0.04$ keV at the 90% confidence level). Therefore, to fit the combined spectra extracted from the 3 annuli between 5" and 60", a proper model has to include at least 3 emission components (1 hard + 2 soft components). In this 3-component spectral fit, the acceptable range of Z is 0.25-1.3 solar (at 90 % confidence).

Previous measurements of the metal abundance of the hot ISM of NGC 1316, from the analysis of ASCA spectra, are reported by Iyomoto et al. (1998) and Matsushita et al. (2000). ASCA-SIS had a spectral resolution similar to that of *Chandra*-ACIS, but a very large (arcmin) beam, so that no spatial discrimination of spectral emission components was possible.

Iyomoto et al. (1998) found a slightly higher temperature (kT = 0.77 ± 0.04 keV), but a much lower abundance (Z = 0.11 ± 0.06 solar) than we find with *Chandra*. Note that the quoted (formal statistical) errors of both kT and Z are extremely small, and exclude the allowed parameter range from our present analysis. To do a direct comparison of our results with those of Iyomoto et al (1998), we have extracted an integrated *Chandra* spectrum from a 1' circular region (including the AGN and detected point sources), to simulate the lack of spatial resolution of ASCA. Using the same model as theirs (Raymond thermal gas + power law with $N_H$ fixed at $1.5 \times 10^{20}$ cm$^{-2}$) and the same energy range with a lower limit at 0.6 keV (ASCA was not sensitive below 0.6 keV), we were able to reproduce their results with the best fit Z=0.1 solar. The spectral fit is statistically good with $\chi^2$=124 with 143 degrees of freedom. We note that the same model over-predicts at the low energies (0.3-0.6 keV) by as much as 50% and the statistics becomes worse, $\chi^2$=147 for 137 degrees of freedom, if estimated in the energy range of 0.3-5.0 keV. This result shows that there are no substantial calibration problems that may affect these comparisons. Our different conclusions are based on the unequivocal knowledge of spectral complexity that we can derive thanks to the spatial resolution of Chandra, which is guiding our choice of spectral models. Moreover, our ISM data, while containing some contamination from undetected LMXBs, are by and large the cleanest obtainable. Our results are consistent with the more recent conclusions (again from ASCA data) by Matsushita et al. (2000) who showed that more complex models could produce near solar abundances. We do not address the issue of non-solar abundance ratios for Fe and alpha elements, and of the uncertainties in the atomic spectral codes; these were suggested by Matsushita et al. (2000) as possible ways of obtaining Fe abundances closer to solar (as expected from the stellar metallicities, e.g. Arimoto et al 1997). Given the comparable spectral resolution, their considerations still apply to all X-ray CCD data.

## 2.4.5 Spectrum of the `Northern Blob'

As described in section 2.2, an off-center, extended blob of X-ray emission was found at 30-60" north from the nucleus (Figure 2). Its X-ray emission appears to be softer than the surrounding region (see Figure 4). The X-ray colors confirm this softness. In Figure 7, we compare the X-ray colors from the northern blob (red square) and those from the annulus at the similar radius (blue triangle), r = 30-60", after excluding both the northern blob and detected point sources. The difference in the X-ray colors is consistent with the fact that the X-ray emission from the blob is softer. We discuss the possible nature of this off-center extended blob in section 5.3.

### 2.4.6 NGC 1317

As reported in Section 2.1, 3 sources are associated with NGC 1317: an extended source centered on the galaxy, and two point-like sources within the D25 ellipse (Table 1). The X-ray spectrum of NGC 1317 was extracted within a 20 arcsec radius. Although the spectral fitting (Table 4) does not constrain well all the spectral parameters, it suggests that the X-ray emission is soft (0.3-0.4 keV). The soft X-ray spectrum is consistent with thermal emission from a hot ISM. Given the poor counting statistics, we cannot comment on a LMXB contribution to the extended emission. The X-ray colors of the central source of NGC 1317 are consistent with the spectral fitting results, suggesting soft X-ray emission with $C21=0.27 \pm 0.03$ and $C32=0.83 \pm 0.03$, which would place this source near (slightly above) the blue triangle in Figure 7. The second brightest source (CXUO J032246.1-370552) appears to have a relatively hard spectrum (suggesting that it is a LMXB associated with NGC 1317) with $C21=-0.28 \pm 0.2$ and $C32=0.73 \pm 0.1$, which approximately place it in Figure 7 in the middle of point sources found in NGC 1316. Its luminosity ($Lx = 5 \times 10^{38}$ erg sec$^{-1}$, Table 1) is consistent with a luminous LMXB.

```
               Table 4. Spectral Fitting of Diffuse X-ray Emission in NGC 1317
______________________________________________________________________________________

model N_H^+           KT_1        Z         KT_2                 Fx*               χ²_reduced (χ²/dof)
                                                        total 1st-comp 2nd-comp
______________________________________________________________________________________

(r <20" ; total counts = 553)

1   0.00 (0.03) 0.33 (0.04)  1.00                    0.53                  1.00 (= 21.1/21)
2   0.14 (0.02) 0.31 (0.03)  0.04 (0.01)             0.58                  0.37 (=  7.3/20)
3   0.13 (0.02) 0.31 (0.04)  1.00        0.33 (0.04) 0.58  0.28  0.30      0.39 (=  7.3/19)
4   0.00 (0.02) 0.32 (0.03)  1.00        5.00        0.72  0.46  0.27      0.62 (= 12.4/20)
______________________________________________________________________________________

Errors in parentheses are at the 90% confidence level for one interesting parameter.
+ N_H is in unit of 10^22 cm^-2
* Fx is in unit of 10^-13 erg s^-1 cm^-2 in a 0.3 – 8.0 keV energy band

Model 1: one-component model: MEKAL with fixed Z
Model 2: one-component model: MEKAL with varying Z
Model 3: two-component model: MEKAL + Bremsstrahlung
Model 4: two-component model: MEKAL + Bremsstrahlung (with fixed kT_2=5.0)
```

### 3. The XLF and the Hidden Population of Point Sources

To determine the XLF of the point sources in NGC 1316, we have used 66 sources. These are the sources detected in CCD S3 at r > 20", to avoid the region of most intense diffuse emission and possible source confusion. We excluded 4 sources detected in CCD S2, to avoid the effect of different instrumental responses in the determination of flux limits. The cumulative XLF is shown in Figure 10, where both binned (open squares) and unbinned (dotted line) distributions are plotted. The lowest luminosity point corresponds to the lowest luminosity sources detected in NGC 1316. A single power-law model was used to reproduce the data with count rate > 0.6 count ksec$^{-1}$ (corresponding to Lx > 1.2 x $10^{38}$ erg sec$^{-1}$) where completeness starts to drop (see below). Because errors in cumulative counts are not independent, we have applied the maximum likelihood method to fit the unbinned data and the minimum $\chi^2$ method to fit the binned data. The best-fit power-law slopes determined with these two approaches are similar: -1.32 ($\pm$ 0.05) with the minimum $\chi^2$ and -1.25 with the maximum likelihood method. These values are consistent with those found in other early type galaxies (e.g. Sarazin et al. 2000), but much steeper than those in spirals and star burst galaxies (Prestwich et al. 2001; Zezas and Fabbiano 2002).

The XLF becomes flatter with decreasing source count rate below 0.6 count ksec$^{-1}$. We have investigated this effect further to ascertain if it may be due to a physical break in the distribution of source luminosities (e.g. as suggested by Sarazin et al. 2000; Blanton et al. 2001 in other early-type galaxies), or it may result from biases affecting the detection threshold of the data. We have explicitly considered two types of errors that may affect the lower count rate detections, and we have corrected the XLF accordingly: type I error is the probability to find a spurious source and type II error is the probability to miss a real source. Large sets of simulations were performed (using the MARX simulator available in http://space.mit.edu/CXO/MARX) to quantitatively determine type I and II errors. For the determination of type I error, we have also used publicly available, long-exposure archival *Chandra* data to complement the simulation study, by splitting the long exposure data into multiple short ones to test how many false sources are identified in short-exposure data, but not in the long-exposure data. This second approach could in principle be affected by source variability, so a greater "spurious" detection may be expected than when using simulated data. We have found that type I error is always as good as or better than what is expected based on the threshold parameter in **wavdetect**. With the selected threshold of $10^{-6}$, the number of spurious sources per CCD is 1 or less. Instead, type II error (the incompleteness effect) is more complex and varies significantly as a function of source count, background count, off-axis distance, and detection threshold.

We estimated the incompleteness corrections in two different ways (see the Appendix for details): (1) **Forward correction** - we estimated from simulations the detection probabilities for different source counts at different off-axis distances and for a range of background (+ diffuse ISM) emission. This was done for nine galactocentric annuli with radii ranging from 20" to 360". We then added corrections for the radial variation of the number of detected point sources (roughly following a r$^{-1}$ distribution) as determined in section 2.3. Corrections were also applied for the uneven sky coverage within the CCD S3 chip and for telescope vignetting (both included in the exposure map). The appropriate correction factor, including all the above effects, was applied to each bin of the differential XLF, from which a corrected cumulative XLF was then derived. This exercise showed that incompleteness can be substantial: 50% for a typical source of 15 counts with background of 0.05 counts per pixel at 5' off-axis (see Appendix). (2) **Backward correction** - we measured the correction factor, directly applicable to the cumulative XLF, by adding one-by-one a large number of simulated sources to the real image data and determining whether each one could be detected. In both approaches, no assumption was made on the behavior of the XLF below our detection threshold.

In Figure 10, we plot both corrected XLFs: the forward correction is marked by red squares and the backward correction by blue triangles. The two corrections are consistent with each other and clearly demonstrate that the single power law (the solid line histogram), that was determined from the uncorrected XLF at $L_x > 1.2 \times 10^{38}$ erg sec$^{-1}$, does fit well the corrected data. In conclusion, the apparent flattening of the XLF of NGC 1316 is fully accountable by incompleteness and we find no evidence of breaks, down to source luminosities of a ~$3 \times 10^{37}$ erg sec$^{-1}$.

At the high $L_x$ end, as the number of sources decreases rapidly with increasing $L_x$, we can not statistically exclude a possible break at $L_x > 5 \times 10^{38}$ erg sec$^{-1}$, although it is not required by the data. Finoguenov and Jones (2002) reported a XLF break at $5 \times 10^{38}$ erg sec$^{-1}$ in M84, an elliptical galaxy in Virgo cluster, which could be reconciled with the Eddington limit by reducing $L_x$ by a factor of 0.5 by adopting a softer spectrum than the average spectra of point sources. However, we do not see any concentration of point sources with such soft spectra (see Figure 7).

The contribution of the hidden population of LMXBs to the observed X-ray emission depends on the low-luminosity XLF. As seen in the bulge of M31, it is possible to have a break at $L_x \sim 10^{37}$ erg sec$^{-1}$ (Kong et al. 2002). If the XLF continues as an uninterrupted power-law down to $10^{37}$ erg sec$^{-1}$, the total X-ray luminosity of the hidden population of LMXBs of NGC 1316 would be comparable to that of the detected sources. If the XLF has a break at the lowest observed $L_x$ (i.e., ~$3 \times 10^{37}$ erg sec$^{-1}$), the luminosity ratio of hidden to detected sources is about 0.4. As demonstrated in section 2.4.3, the amount of hard component required in the spectral fitting of the diffuse X-ray emission is fully within this acceptable range.

We have applied the same technique of completeness correction to other early type galaxies for which *Chandra* data are publicly available. In general, near the center of early type galaxies where the PSF is usually smallest as the galaxy center is mostly likely at the focus, the strong diffuse emission considerably reduces the detection probability. On the other hand, at the outskirts, the PSF becomes considerably larger, hence the detection probability becomes lower. So far we have found no clear case where a broken power law is required. The results will be reported elsewhere (Kim et al. 2002).

## 4. Summary of Emission Components

The X-ray fluxes and luminosities of the different emission components of NGC 1316 are summarized in Table 5. For the detected LMXBs, all point sources are included except two sources at the center of NGC 1316, which may be associated with features of the ISM (Section 2.1). We estimate that ~10% of LMXBs are outside the field of view in CCD S3, based on the spatial distribution of point sources detected in CCD S3. We have corrected the entries in the table for this effect. We give two estimates of the X-ray emission from the undetected LMXBs, obtained by taking the lower limit of the XLF at either $1 \times 10^{37}$ erg sec$^{-1}$ or $3 \times 10^{37}$ erg sec$^{-1}$. We note that the integrated Lx of the undetected LMXBs could be as large as that of the hot ISM, although their total counts is smaller because of the harder LMXB spectrum. We also note that the error on Lx caused by the uncertainty on the hidden LMXB contribution is much larger than the formal statistical error.

```
                       Table 5. X-ray Luminosities of Individual Components
________________________________________________________________________________________
                                               -13      -2    -1         40        -1
Component                      Net Counts  Fx (10   erg cm   sec  )  Lx (10   erg sec  )
________________________________________________________________________________________

AGN                                  520           1.3                      0.51
LMXB (detected)                     2130           5.8                      2.3
Diffuse (gas + undetected LMXB)     6400           7.8                      3.1

Undetected LMXB                  840-2100          2.3-5.8                  0.92-2.3
Gas only                        4460-5620          5.4-6.8                  2.2-2.7

Total                               9050          16.2-18.3                 6.4-7.3
________________________________________________________________________________________

Notes.
- Fx and Lx are in a 0.3-8 keV energy band.
- ECF = 6.0x10^-15, 6.7x10^-15 and 3.0x10^-15 for 1 count ksec^-1 are used for AGN, LMXB
  and gas components, respectively. For the diffuse emission (gas + undetected LMXBs),
  ECF for gas was used.
- About 10% of LMXBs are added because of the limited field of view.
- The X-ray emission from the undetected LMXBs is estimated by assuming the lower limit
  of XLF at 1 - 3 x 10^37 erg sec^-1.
________________________________________________________________________________________
```

## 5. Discussion

The *Chandra* observations of NGC 1316 have bearing on several questions. NGC 1316 is an X-ray-faint elliptical galaxy, and therefore our results have direct relevance for our understanding of the nature of the X-ray emission of this type of galaxies (see Fabbiano 1989). NGC 1316 is also the Fornax A radio galaxy, and therefore we can explore the interaction of the radio jets with the hot ISM in this system, and we can study the X-ray properties of the active nucleus. Finally, NGC 1316 is the likely product of a merger (eg., Schweizer 1980) and therefore our results are relevant for understanding merger evolution. In what follows we will address these three main topics in turn.

### 5.1 NGC 1316 as an X-ray-faint Elliptical galaxy

While hot ISM was discovered over 20 years ago in E and S0 galaxies (Forman et al. 1979), the properties of this medium and its evolution have been the subject of heated discussion (see e.g., Fabbiano 1989; Ciotti et al. 1991; Eskridge et al. 1995a, b; Arimoto et al 1997). X-ray-faint E and S0 galaxies, of which NGC 1316 is one, were particularly debated. These are galaxies with X-ray to optical luminosity ratios 100-1000 times smaller than those of hot-halo dominated Virgo galaxies. A question that was raised early on is the role of the hot ISM in the overall X-ray emission of these galaxies versus that of LMXBs (ISM dominant: e.g., Forman, Jones & Tucker 1985; LMXBs dominant: Trinchieri & Fabbiano 1985; Canizares, Fabbiano & Trinchieri 1987; both present: Fabbiano, Kim & Trinchieri 1994; Pellegrini & Fabbiano 1994). With the advent of CCD X-ray spectrometers (ASCA SIS), it became clear that the X-ray spectra were complex, with a thermal hot ISM component and a harder component, possibly from LMXBs (Matsushita et al. 1994). Still, the lack of spatial resolutions made the interpretation of these data ambiguous at times: in particular a low-luminosity AGN could also contribute to the hard component; LMXBs were suggested to contribute to the soft emission (Irwin & Sarazin 1998). Moreover, the issue of the metal abundance of this hot ISM remains controversial and open (see Arimoto et al 1997; Matsushita et al 2000), and ASCA could not establish the spatial/spectral properties of this medium, which are important for understanding its physical state (e.g. Pellegrini & Ciotti 1998).

NGC 1316 joins a growing body of E and S0 galaxies observed with *Chandra* at sub-arcsecond resolution. With these data we now have uncontroversial proof of the existence of populations of X-ray binaries in all E and S0 galaxies and of their large contribution to the emission of X-ray-faint galaxies (see Sarazin et al. 2000; Sarazin et al. 2001; Blanton et al. 2001; Angelini et al. 2001). In NGC 1316, we detect 81 non-nuclear point-sources, with a surface brightness distribution following that of the optical stellar light. These sources have the hard (kT ~ 5 – 6 keV) spectrum characteristic of LMXBs, and luminosities in the range $2 \times 10^{37} - 2 \times 10^{39}$ erg sec$^{-1}$. ~35% of these sources (with one ULX) have $L_x > 2 \times 10^{38}$ erg sec$^{-1}$, the Eddington luminosity of neutron star X-ray binaries. One source is a supersoft X-ray source (SSS). Five sources have Globular Cluster (GC) counterparts (Table 1); sources in GC are found frequently in ellipticals, and GC have been suggested as the birthplace for a large fraction of LMXBs in these galaxies (White et al 2002).

In contrast with earlier reports for other early-type galaxies (e.g., Sarazin et al. 2000), we find that the X-ray luminosity function (XLF) of the NGC 1316 sources does not have a strong break at the Eddington luminosity of an accreting neutron star, but appears to follow an uninterrupted ~1.3 steep power-law down to our detection threshold luminosity of ~$3 \times 10^{37}$ erg sec$^{-1}$, once incompleteness corrections are applied. The presence of a substantial amount of faint LMXBs is also suggested by the spectra of the diffuse emission within a 5" – 60" annulus: after removal of all the detected point sources in this region, they still contain a hard component, presumably from undetected LMXBs, accounting for ~25% of the emission. Given that the emission at the outer radii is dominated by the point-like component in NGC 1316 (Section 2.3), the total LMXB contribution to the emission of this galaxy is larger. Extrapolating the XLF at lower luminosities, the total (detected and undetected) LMXB emission could account for ~1/2 to ~2/3 of the total X-ray luminosity of NGC 1316, depending on the low-luminosity break of the XLF (Table 5). The larger estimate was calculated for a XLF break at 1 x $10^{37}$ erg sec$^{-1}$, analogous to that reported in the bulge of M31 by Kong et al (2002).

The *Chandra* data also show hot ISM in NGC 1316, confirming earlier ROSAT results (KMF). This hot ISM is more centrally concentrated than the LMXB and stellar light distributions, and appears to be cooler at larger radii. The steep X-ray surface brightness profile of the diffuse emission, the average temperature of the hot ISM (kT ~ 0.5 - 0.6 keV) and the radially decreasing temperature profiles we derive from our spectral analysis, together with the low X-ray luminosity (~2.5 x $10^{40}$ erg sec$^{-1}$), could be consistent with the Pellegrini & Ciotti (1998) picture of partial wind flows for X-ray faint ellipticals.

The abundance of heavy elements has been identified as one of the key quantities for our understanding of the evolutionary status of the hot ISM in terms of the supernova rate (Ia and II), IMF, mass ejection and onset of galactic winds (eg., Renzini et al. 1993, Matteucci and Gibson 1995). Yet its measurement is difficult. While the ACIS spectral resolution is comparable to that of the ASCA and XMM CCD detectors, the spatial resolution of *Chandra* can remove some of the ambiguity in the choice of spectral models for spatially/spectrally complex sources, such as galaxies. Different models may all give acceptable spectral fits, and the accepted practice is to adopt the simplest possible model; however the answer may not be astrophysically meaningful, because of the complexity of the sources we're attempting to fit. Fits of the ROSAT and ASCA data with single temperature Raymond-Smith spectra (+ hard component) suggested a hot ISM almost totally devoid of metals in X-ray faint galaxies (eg., Awaki et al. 1994; Loewenstein et al. 1994; Davis and White 1996; Matsumoto et al. 1997; Iyomoto et al 1998 for NGC 1316), while more complex models allowed a metal content more in keeping with the stellar metallicities (e.g., Trinchieri et al. 1994; Kim et al., 1996; Buote and Fabian, 1998; Matsushita et al 2000). As shown in Section 2.4.4, we can remove some of this ambiguity in NGC 1316 by excluding the regions of the data-cube occupied by detected point-like sources (AGN and LMXBs) from the spectral data and establishing the existence of variations (in ~10" radial scales) of the temperature of the hot ISM. While we do not address uncertainties in the spectral codes, that persist in all the spectral analysis of thermal X-ray emission (see Matsushita et al 2000), our analysis, guided by the high resolution images of Chandra, convincingly excludes the possibility of extremely low metal abundances in the hot ISM (see Iyomoto et al 1998), giving an acceptable (90% confidence) range of Z = 0.25 – 1.3 solar. However, we note that the Fe abundance is still lower than what might be expected (e.g., Arimoto et al. 1997), particularly if the partial wind is driven by SN Ia.

A final topic of considerable debate has been the ability of measuring the gravitational mass of E and S0 galaxies, based on the assumption of hot halos in hydrostatic equilibrium (see e.g. Fabbiano 1989; Trinchieri, Fabbiano & Canizares 1986). In particular, it was pointed out by the above authors that this method would not apply (or at least would be extremely uncertain) in the case of X-ray-faint galaxies. This is certainly true in the case of NGC 1316, for which a mass of 2.0 x $10^{12}$ solar masses was estimated, based on the *Einstein Observatory* data (Forman, Jones & Tucker 1985).

The mass measurement based on the equation of hydrostatic equilibrium (see e.g. discussion in Trinchieri et al 1986) hinges on 4 quantities (outer radius of the halo, temperature, density and temperature gradients at this radius). In NGC 1316, we have found that the X-ray binaries dominate the emission at the outer radii, while the gas extends out to ~ 60'' (5.4 kpc for the distance adopted in this paper). Taking into account the different distance used by Forman et al, the radius of the halo used in the mass estimate was ~19 kpc, comparable to the radius of the region within which LMXBs are detected. The temperature of the halo was assumed to be ~1 keV (we now know it is 0.5 keV). Clearly, data with less angular and spectral resolution than *Chandra*'s would not (1) allow to separate point sources from diffuse emission, giving a mistaken larger radius for the hot halo, and a more relaxed radial distribution; (2) also the temperature of the halo may appear larger than it really is, because of contamination with the harder LMXB emission, if these cannot be removed spatially or spectrally (the CCD resolution of ASCA helped on the spectral analysis for the first time); (3) the halo may not be in a state of hydrostatic equilibrium, if for example is subject to winds at the outer radii, as it may be the case, based on the present data. This type of uncertainties is common to all pre-*Chandra* mass measurements of X-ray faint E and S0 galaxies: because of the problems exemplified in the case of NGC 1316, these measurements would tend to err in excess, even if the hot ISM were in hydrostatic equilibrium. Caution should be exercised even when interpreting XMM data, although the XMM moderate spatial resolution and CCD spectral capabilities allow a better discrimination than any of the pre-*Chandra* missions.

**5.2. The Fornax A radio galaxy**

Figure 3 shows the direction of radio jets (from the 1.5GHz VLA map of Figure 2b in Geldzahler and Fomalont 1984) superposed on the X-ray image. The radio jets propagate along PA ~ 120° and 320°, and slightly bend at their ends toward PA ~ 90° and 270°, respectively (Geldzahler and Fomalont 1984). It is clearly seen that the radio jets in projection (1) run perpendicular to the direction of the sub-arcmin scale NE-SW elongation of the central X-ray distribution, which is also the direction of the optical figure, and (2) propagate through the valleys of the X-ray emission, confirming the results of the ROSAT HRI observations (KFM). This alignment suggests that the X-ray valleys are cavities in the hot ISM caused by exclusion of hot gas from the volumes occupied by the radio jets. Similar relations between the radio jets/lobes and X-ray features have been reported in the *Chandra* observations of NGC 4374 (Finoguenov and Jones 2001), Hydra A (McNamara et al. 2000), Perseus cluster, (Fabian et al. 2000) and Abell 4095 (Heinz et al., 2002).

In particular, the hourglass-like feature seen in the core of Abell 4095, is similar to the X-ray feature in NGC 1316, although only one of (possibly) two X-ray cavities in Abell 4095 coincides with the radio emission. Based on the hydrodynamic simulations of Reynolds et al. (2001), Heinz et al. (2002) have interpreted the hourglass feature in Abell 4095 as a sideways expansion of the jet's surface caused by a compression wave (or sonic boom) which would follow after driving a strong shock. However, in NGC 1316 the jets appear to escape the denser halo. It is, therefore, not clear whether a sonic boom could explain the NE-SW elongation. This elongation could simply reflect the hot ISM sitting on the local, spherically asymmetric potential because it also follows the major axis of the galaxy.

Although NGC 1316 has luminous radio lobes, the radio emission from the active galactic nucleus (AGN) is relatively faint. Fornax A contains a relatively weak radio core with a total power of $\sim 10^{38}$ erg sec$^{-1}$ (Geldzahler and Fomalont 1984). We note that the current nuclear radio power is too weak to sweep the hot ISM out of the X-ray cavities as it would take longer than $\sim 10$ Gyr for the minimum energy required by pV $\sim 10^{56}$ erg, where the pressure was taken from KFM. This indicates that the nuclear activity was much stronger in the past, as suggested by Ekers et al. (1983). Mergers, which took place a few Gyr ago, might have ignited the AGN activity, which in turn swept out the hot gas and also powered the huge radio lobes.

The AGN in NGC 1316 has a LINER-type spectrum (Veron-Cetti and Veron 1986; Baum, Heckman and van Breugel 1992) and HST observations have revealed a nuclear UV-bright point source (Fabbiano, Fassnacht and Trinchieri 1994b). With our *Chandra* ACIS observation we have discovered the X-ray counterpart of this nuclear emission, which is prominent in the hard band image (Figure 4). Also the nuclear component is required to fit the radial profile (Figure 6) and the X-ray spectrum of the nuclear region (Figure 8). For this source we have measured a flux Fx = 1.3 x 10$^{-13}$ erg s$^{-1}$ cm$^{-2}$ (in 0.3-8 keV) and a luminosity Lx = 5 x 10$^{39}$ erg s$^{-1}$ (see Section 2.2 and 2.4.2), consistent with the upper limit estimated by KFM. The spectrum of this source is well fitted with a power-law with photon index $\Gamma \sim 1.7$. Based on the relationship between $M_{BH}$ and the stellar velocity dispersion σ (Tremaine et al. 2002), the estimated mass of the nuclear black hole of NGC 1316 is about $M_{BH}$ = 1-2 x 10$^8$ M⊙ for σ = 221 km sec$^{-1}$ (Kuntschner 2000). The X-ray luminosity of the nuclear source then corresponds only to 10$^{-6}$ x $L_{Eddington}$. Both luminosity and spectral characteristics are reminiscent of those of the X-ray faint nucleus of IC 1459 reported in Fabbiano et al. (2002) to which we refer for further discussions of this type of source (see also Ho et al. 2001, Pellegrini et al. 2002).

We are conducting multi-frequency VLA observations to obtain high S/N radio images of the nucleus and jets with a resolution comparable to that of the *Chandra* observations. We will report in-depth comparison between the X-ray and radio data and their implications in another paper.

## 5.3 Features and history of the hot ISM

NGC 1316 is an X-ray faint elliptical galaxy, i.e. it has relatively little hot ISM, much less than would be expected from steady accumulation of the gas shed by evolving stars during an undisturbed galaxy lifetime (see Fabbiano 1989, Eskridge, Fabbiano and Kim, 1995a, b). Since NGC 1316 is the likely product of recent mergers (Schweizer 1980), this lack of hot ISM could be related to its recent merging history. KMF noted that the total amount of the hot ISM (~$10^9$ M⊙) is comparable with what would have been accumulated for a few Gyr. Other merger remnant galaxies have also been found to be X-ray faint (Fabbiano & Schweizer 1995; Hibbard et al. 2002). A close look at a nearby merger remnant, the X-ray faint S0 NGC 5128 (Cen A), with *Chandra*, has revealed a complex distribution of cold and hot ISM (Karovska et al. 2002). Similarly, in NGC 1316 the *Chandra* images confirm the complex distribution of the hot ISM suggested by ROSAT (KFM). Besides the cavities created by the expanding radio jets that we have just discussed, the hot ISM presents other fine structures that may be indicative either of absorption by cold dusty ISM in the line of sight but within the parent galaxy, or of a multi-phase ISM, where hot and cold phases are intermingled. This is shown by Figure 11 where we have compared the *Chandra* X-ray surface brightness distribution with the archival HST WFPC2 image taken with a F450W filter (central wavelength at 4520A and bandwidth of 958A). Many prominent dust features are visible in the HST WFPC2 image and we indicate them with green lines in the figure. They correspond to the central part of the lower resolution dust lane map of Schweizer (1980). Some dust features are seen where the X-ray surface brightness is reduced, suggesting absorption.

As discussed in Section 2.2, we have found an off-center, 2-3 kpc extended blob of X-ray emission at 30-60" north from the nucleus, which is also cooler than the surrounding hot ISM (section 2.2 and 2.4.3). We estimate a cooling time of about 2Gyr for this feature. This cooling time is similar to the time of the latest merger event in NGC 1316 (Schweizer 1980; Ekers et al. 1983; confirmed by more recent optical studies of star clusters, Goudfrooij et al. 2001). It is, therefore, possible that the formation of the blob may be related to compression of the ISM during the last merging of NGC 1316. A similar off-center blob was seen in the *Chandra* observations of NGC 507. Given its position relative to the radio lobe, Forman et al. (2001) have interpreted it as being ISM displaced by the radio lobe. However, as the direction to the blob in NGC 1316 is almost perpendicular to that of the ratio jets/lobes, the cooling blob in NGC 1316 appears to be unrelated to the radio jets.

The origin of the various phases of the ISM in early type galaxies has been discussed since the early detection of cold/warm ISM in forms of HI, far IR, CO, dust lanes, and Hα (eg., Jura et al. 1987; Kim 1989). The two options that have been advanced are that the cold/warm ISM could be either of external origin, accreted by way of merging and close encounter episodes with other galaxies, or that it could be the byproduct of cooling hot ISM originating from internal stellar mass loss, e.g. via cooling flows (eg. Knapp et al. 1985; Kim et al. 1988). In NGC 1316, the kinematics of the ionized (Schweizer 1980) and molecular gas (Sage and Galletta 1993), in comparison to those of the stellar system (Bosma, Smith and Wellington 1985), indicate that cold/warm ISM and stars are decoupled, suggesting an external origin for this phase of the ISM. This is not surprising, given the likely merging history of NGC 1316. Instead, since the hot ISM is likely to originate from stellar evolution (see review in Fabbiano 1989), we would expect it to share the stellar kinematics and to be decoupled from the cold ISM. This possibility was suggested by earlier works. With the ROSAT HRI data, we pointed out that the morphology of the larger scale distribution of the hot ISM suggests a disk co-aligned with the stellar system central X-ray flattening (KFM), and that the hot and cold ISM are not morphologically related (Mackie and Fabbiano 1998). A future deep *Chandra* observation of NGC 1316 would give us the means to explore this point further by examining in detail the properties of the interface between cold and hot ISM.

**6. Conclusions**

The high resolution *Chandra* ACIS observations of the X-ray-faint merger remnant NGC 1316, the Fornax A radio galaxy, have provided an unprecedented look at this complex system. In summary:

1. We have confirmed the presence of cavities in the hot ISM related to the expanding radio jets, first suggested on the basis of a ROSAT HRI observation (KFM).

2. We have discovered a low-luminosity AGN (Fx = 1.3 x $10^{-13}$ or Lx = 5 x $10^{39}$ erg/s). Although the nuclear source is weak, its power-law spectrum is typical of luminous AGN, with a best-fit photon index of 1.7.

3. We have found 81 point sources in NGC 1316 and 3 in the nearby galaxy NGC 1317. Their average spectrum is hard (~ 5keV) and the radial distribution follows that of the optical light, indicating that most of them are LMXBs associated with the galaxy. Five of these sources have GC counterparts and one is supersoft. 35% of them (with one ULX) have X-ray luminosities in excess of the Eddington luminosity of an accreting neutron star.

4. The X-ray luminosity function (XLF) of the NGC 1316 sources is well represented by a single power law with a slope of -1.3, which is consistent with those of other early type galaxies studied with *Chandra* (Sarazin et al. 2000) but much steeper than that of spiral or starburst galaxies (Prestwitch et al. 2001). The uncorrected XLF shows a flattening at the faint end, which is removed once corrections are applied for the incompleteness errors stemming from the local diffuse emission at small galactocentric radii and from the increasing PSFs at larger off-axis angles. Once these corrections are applied, the XLF follow an uninterrupted power-law, with no breaks, down to a luminosity of ~ 3 x $10^{37}$ erg $sec^{-1}$, indicating that there are still a considerable number of faint, undetected sources. Even without the incompleteness correction, the single power law is still applicable to the observed XLF down to $10^{38}$ erg $sec^{-1}$, at luminosities below the break at $L_{Eddington}$ of neutron star binaries reported in other X-ray faint elliptical XLF (Sarazin et al. 2000; Blanton et al. 2001).

5. The radial profile of the diffuse X-ray emission varies at different position angles: it is more steeper toward E-W than N-S. Regardless of PAs, the radial profile of the diffuse X-ray emission is much steeper than that of the optical light unlikely that of the LMXBs (see point 3 above). This means that the relative contribution of the LMXBs to the diffuse emission increases with galactocentric radius. At large radii, the diffuse emission is likely to be dominated by unresolved LMXBs rather than by the hot ISM.

6. Spectrally, the diffuse X-ray emission (extracted from narrow annuli and after subtracting all detected point sources) is well reproduced by two component models: in the nuclear region, thermal optically thin gaseous emission plus an AGN power-law component; at larger radii, gaseous emission plus a hard component from the undetected LMXBs. The best fit temperature for the gaseous component is kT ~ 0.5-0.6 keV, with a small but statistically significant decrease (by 0.12 keV) from the center to 60". We constrain the metal abundance of the hot ISM to be in the range 0.25 – 1.3 solar (90% confidence).

7. The hot ISM presents a number of features that may be related both to interaction with the radio emission (the cavities) and to the recent merging history of NGC 1316 (Schweizer 1980). In particular, there is clear intermingling of cold and hot ISM in the central regions. We have also found a well-defined soft emission region (blob) at ~40" north from the nucleus. Given its cooling time of 2Gyr, it could be a cooler blob falling into the center, formed at or after the recent merger. Overall, the relatively small amount of hot ISM in NGC 1316 is consistent with the observations of other recent mergers (Fabbiano & Schweizer 1995; Hibbard et al. 2002). The current radio power of the nucleus and jets is too weak to sweep out the hot gas out of the cavities, indicating that the AGN was more powerful in the past, around the time of the recent merger. The irregular morphology of this ISM and the interplay between hot and cold phases are also reminiscent of the recent *Chandra* results on NGC 5128 (Cen A) another nearby merger remnant (Karovska et al 2002).

These results (particularly points 2-6) have an important bearing on the interpretation of lower angular resolution data of early-type galaxies, both pre-*Chandra*, XMM and of future planned missions (Constellation X; XEUS). It is only with sub-arcsecond resolution that we can resolve the LMXB population and the low luminosity nuclear AGN. This, coupled with the wide energy band and spectral resolution of ACIS, leads to a clear definition of the spatial and spectral properties of the different emission components. As a result, we can conclude with confidence that: (1) LMXBs are a dominant contributor in the X-ray faint early-type NGC 1316 (see also Sarazin et al. 2000; Blanton et al 2001 for other examples of X-ray-faint early-type galaxies). These results confirm the early inference based on Einstein and ROSAT observations (see Fabbiano 1989; Trinchieri & Fabbiano 1985; Canizares, Fabbiano & Trinchieri 1987). (2) ROSAT and ASCA reports of an ISM devoid of metals in these galaxies were biased by the lack of a proper appreciation of the spectral complexity of these systems (as discussed by Matsushita et al 2000), which becomes evident when a spatially resolved spectral analysis can be undertaken. (3) X-ray-based mass measurements for these systems (see Fabbiano 1989) that do not take into account the real distribution of the hot ISM (both spatially and spectrally, for which adequate spatial and spectral resolution are needed) are bound to be wrong and should not be attempted.

This work was supported by NASA grant GO1-2082X. We thank the CXC DS and SDS teams for their supports in pipeline processing and data analysis. We thank D. Harris for

**Appendix: Incompleteness Correction.**

A. Forward Correction

The detection probability (type II error) is a complex function of background count per pixel, different source counts and off-axis distances. Large sets of simulations based on MARX were performed (Kashyap et al. 2002) to establish the behavior of the detection probability in a wide parameter space. The results are illustrated in Figure 12: the detection probability decreases as the *effective background* rate (which in our case includes both field background and galaxian diffuse emission) and the off-axis angle increase and as the source count decreases. Incompleteness could be as large as 50% for a typical source of 15 counts with background of 0.05 counts per pixel at 5' off-axis. In the *Chandra* observations of NGC 1316, the effective background ranges from 0.2 count per pixel at 20" from the center of NGC 1316 to 0.03 count per pixel at 60" and beyond. This is large enough to result in a significant amount of type II errors for weak sources.

For the source counts corresponding to the three lowest luminosity bins of Figure 10, we have determined incompleteness-correction factors, applicable to the **differential XLF**, as summarized in Table A-1.

```
                         Table A-1.                              .
              Correction factors for the last 3 bins of XLF            .
________________________________________________________________________

r_in   r_out  bkg_rate  area       F1     F2     F3         det_prob
(")    (")    cnt/pix   arcsec^2   note1  note2  note3   bin1   bin2   bin3
________________________________________________________________________

 20     30    0.12       1570     1000    1.0    716    0.7    0.6    0.1
 30     40    0.09       2198      714    1.0    717    0.75   0.65   0.15
 40     50    0.07       2826      555    1.0    718    0.75   0.65   0.15
 50     60    0.06       3454      455    1.0    714    0.75   0.65   0.15
 60    120    0.04      33912      250    1.0    714    0.8    0.7    0.2
120    180    0.03      56520      156    1.0    591    0.6    0.5    0.1
180    240    0.03      79128      114    0.94   415    0.3    0.2    0
240    300    0.03     101736       89    0.79   388    0.1    0      0
300    360    0.03     124344       74    0.70   334    0.1    0      0
________________________________________________________________________

Weighted Sum*            26897**                        14261  11571  2611
correction factor           1.0                          0.53   0.43  0.09
________________________________________________________________________

Note1. Number of point sources in unit area (arbitrarily scaled).
Note2. Area coverage within CCD S3 (arbitrarily scaled).
Note3. Telescope vignetting (arbitrarily scaled).
bin1.  log (cnt rate) = -3.3 ; counts = 9.8 – 15.6
bin2.  log (cnt rate) = -3.5 ; counts = 6.2 –  9.8
bin3.  log (cnt rate) = -3.7 ; counts = 3.9 –  6.2
*      Weighted sum is Σ(det_prob*area*F1*F2*F3).
**     Weighted sum with det_prob=1.0
```

Detection probabilities were calculated for nine galactocentric radial bins (from 20" to 360"), appropriate for the bin's off-axis angle and effective background emission. These detections probabilities are listed in the last 3 columns of Table A-1. Then we applied three factors to correct for (1) that the number of point sources in unit area decreases as the distance from the galactic center increases (F1 in Table A-1), (2) that the outer part of the galaxy is not fully covered by CCD S3 (F2 in Table A-1) and (3) finally that the telescope vignetting and CCD response vary at different positions (F3 in Table A-1). Weighted-mean correction factors for each luminosity bin were then determined by multiplying these factors in each radial bin, summing them together and normalizing by the value similarly determined with detection probability equal to 1.0. The estimated correction factor in each bin is then applied to the differential XLF, and the cumulative XLF is derived from the corrected differential XLF. The 3 red squares in Figure 10 indicate the cumulative XLF with these corrections.

The radial dependence of the detection probabilities in Table A-1 reflects the effects of the variation in background level and PSF. We have excluded r < 20" where the background is larger because of the more intense diffuse emission, and also source confusion may be an issue, because of the increased surface density of point sources. The detection probabilities increase with radius and peak between 60"-120", because of the decrease of the background (+diffuse) emission. At larger radii, the predominant effect is the increase of the PSF, which results in a significant decrease of the detection probability.

B. Backward Correction

To confirm the incompleteness corrections, we have also run a simulation, based on the real, observed data of NGC 1316. This time we have determined the correction factor, directly applicable to the **cumulative XLF**, by adding a simulated source to the real image and determining whether it could be detected. To do this, we have selected an input source with a random flux based on a single power law XLF and with a random location based on the optical light distribution of NGC 1316. Note that the assumed XLF is not critical here, because we are only using the ratio of input and output numbers of sources (see Vikhlinin et al. 1995). We have then simulated the sources with MARX**,** added them one-by-one to the real image, and tried to detect them with **wavdetect**. As the real data is being used, those factors necessary in the forward correction (such as radial variations, sky coverage etc.) are already taken into account and the ratio of input and output XLF can be directly used to correct the cumulative XLF. Figures 13-15 show the results of this simulation with 20,000 input sources. In Figure 13, we compare detected counts with input counts. Note that the relative difference increases as the count decreases and the Eddington bias (detected counts tend to be larger than input counts because there are more fainter sources) is present at low counts (see also Vikhlinin et al 1995; Kenter and Murray 2002). Because of the confusion with the strong diffuse emission and with other sources near the center of the galaxy, the detected counts are sometimes considerably different from the input counts. As with the real data analysis in section 2, we do not use sources at r < 20". Figure 14 shows the spatial distribution of simulated sources (green x's). A red circle on a green x indicates a detection. Only 1000 random sources are plotted for visibility. This clearly demonstrates that many faint sources are undetected. Figure 15 exhibits the comparison of input (blue solid line) and measured (red dotted line) XLFs. It is very interesting to note that the measured XLF closely resembles the observed XLF in Figure 10, indicating that the flattening is not real. The ratio of detected to input numbers of sources is applied to correct XLF in Figure 10 (a red triangle). The two results of simulations are fully consistent and reconfirm the amount of incompleteness correction and the validity of an unbroken, single power law down to a few x $10^{37}$ erg sec$^{-1}$.

Table 1. Source List

| source | ccd no. | detect no. | RA (J2000) | DEC (J2000) | source radius (") | net counts | error | counts/ksec | Lx ($10^{38}$ erg s$^{-1}$) | Note |
|---|---|---|---|---|---|---|---|---|---|---|
| <Within the D25 ellipse of NGC 1316> | | | | | | | | | | |
| CXOU J032219.5-371338 | 6 | 6  | 3 22 19.5 | -37 13 37.6 | 3.44 | 11.32  | 4.72  | 0.40  | 1.62  | |
| CXOU J032219.9-371322 | 6 | 5  | 3 22 19.9 | -37 13 21.8 | 3.22 | 6.34   | 3.97  | 0.22  | 0.89  | |
| CXOU J032228.9-371443 | 6 | 1  | 3 22 28.9 | -37 14 42.7 | 3.00 | 3.87   | 3.41  | 0.15  | 0.61  | |
| CXOU J032229.5-371511 | 6 | 4  | 3 22 29.5 | -37 15 11.3 | 3.00 | 8.69   | 4.29  | 0.32  | 1.29  | |
| CXOU J032225.6-371145 | 7 | 83 | 3 22 25.6 | -37 11 45.2 | 3.00 | 8.75   | 4.44  | 0.36  | 0.90  | |
| CXOU J032227.5-371223 | 7 | 78 | 3 22 27.5 | -37 12 23.3 | 3.00 | 7.87   | 4.29  | 0.32  | 0.81  | |
| CXOU J032228.9-371242 | 7 | 63 | 3 22 28.9 | -37 12 41.6 | 3.00 | 22.48  | 6.09  | 0.93  | 2.31  | |
| CXOU J032229.1-371023 | 7 | 62 | 3 22 29.1 | -37 10 23.1 | 3.00 | 4.92   | 3.80  | 0.22  | 0.55  | |
| CXOU J032230.4-371025 | 7 | 61 | 3 22 30.4 | -37 10 25.1 | 3.00 | 7.26   | 4.30  | 0.32  | 0.81  | |
| CXOU J032232.6-371309 | 7 | 60 | 3 22 32.6 | -37 13 9.4  | 3.00 | 5.72   | 3.98  | 0.23  | 0.57  | |
| CXOU J032235.2-371127 | 7 | 26 | 3 22 35.2 | -37 11 26.8 | 3.00 | 32.78  | 7.16  | 1.37  | 3.43  | |
| CXOU J032235.2-371205 | 7 | 59 | 3 22 35.2 | -37 12 5.3  | 3.00 | 10.90  | 4.99  | 0.46  | 1.16  | |
| CXOU J032235.3-371248 | 7 | 25 | 3 22 35.3 | -37 12 47.5 | 3.00 | 26.80  | 6.56  | 1.09  | 2.72  | 02 |
| CXOU J032235.5-371313 | 7 | 24 | 3 22 35.5 | -37 13 13.2 | 3.00 | 75.54  | 9.94  | 3.06  | 7.66  | |
| CXOU J032235.6-371248 | 7 | 58 | 3 22 35.6 | -37 12 47.6 | 3.00 | 23.96  | 6.28  | 0.97  | 2.42  | 02 |
| CXOU J032235.7-371218 | 7 | 57 | 3 22 35.7 | -37 12 18.1 | 3.00 | 11.10  | 5.12  | 0.47  | 1.18  | |
| CXOU J032235.9-371135 | 7 | 23 | 3 22 35.9 | -37 11 34.9 | 3.00 | 12.04  | 5.12  | 0.51  | 1.27  | 09 |
| CXOU J032235.9-371256 | 7 | 56 | 3 22 35.9 | -37 12 56.1 | 3.00 | 8.67   | 4.60  | 0.35  | 0.86  | |
| CXOU J032236.4-370842 | 7 | 22 | 3 22 36.4 | -37 8 42.0  | 5.24 | 11.89  | 5.48  | 0.59  | 1.47  | |
| CXOU J032236.4-371324 | 7 | 82 | 3 22 36.4 | -37 13 23.7 | 3.00 | 72.40  | 9.77  | 2.94  | 7.35  | |
| CXOU J032236.6-371054 | 7 | 54 | 3 22 36.6 | -37 10 54.5 | 3.00 | 3.77   | 3.81  | 0.15  | 0.39  | 05 |
| CXOU J032236.6-371306 | 7 | 55 | 3 22 36.6 | -37 13 6.0  | 3.00 | 17.24  | 5.57  | 0.70  | 1.74  | |
| CXOU J032237.2-371212 | 7 | 21 | 3 22 37.2 | -37 12 11.6 | 3.00 | 7.53   | 4.88  | 0.31  | 0.77  | |
| CXOU J032237.4-371301 | 7 | 53 | 3 22 37.4 | -37 13 0.9  | 3.00 | 7.76   | 4.60  | 0.33  | 0.83  | |
| CXOU J032237.5-371428 | 7 | 52 | 3 22 37.5 | -37 14 28.2 | 3.00 | 7.18   | 4.14  | 0.29  | 0.73  | |
| CXOU J032237.6-371252 | 7 | 51 | 3 22 37.6 | -37 12 51.5 | 3.00 | 20.72  | 6.20  | 0.89  | 2.22  | |
| CXOU J032238.7-371221 | 7 | 20 | 3 22 38.7 | -37 12 20.9 | 3.00 | 4.66   | 4.61  | 0.19  | 0.48  | |
| CXOU J032238.7-371422 | 7 | 50 | 3 22 38.7 | -37 14 21.5 | 3.00 | 31.72  | 6.91  | 1.29  | 3.23  | |
| CXOU J032239.1-371148 | 7 | 19 | 3 22 39.1 | -37 11 48.4 | 3.00 | 23.71  | 7.02  | 1.00  | 2.50  | |
| CXOU J032239.3-371313 | 7 | 18 | 3 22 39.3 | -37 13 12.8 | 3.00 | 19.95  | 6.40  | 0.85  | 2.13  | |
| CXOU J032239.4-371252 | 7 | 48 | 3 22 39.4 | -37 12 52.5 | 3.00 | 8.68   | 5.50  | 0.35  | 0.88  | 02 |
| CXOU J032239.4-371256 | 7 | 49 | 3 22 39.4 | -37 12 56.0 | 3.00 | 10.75  | 5.72  | 0.44  | 1.10  | 02 |
| CXOU J032239.6-371243 | 7 | 76 | 3 22 39.6 | -37 12 43.0 | 3.00 | 22.06  | 7.12  | 0.92  | 2.30  | 05 |
| CXOU J032239.9-371056 | 7 | 47 | 3 22 39.9 | -37 10 55.9 | 3.00 | 22.02  | 6.29  | 0.91  | 2.28  | |
| CXOU J032240.1-371310 | 7 | 17 | 3 22 40.1 | -37 13 10.2 | 3.00 | 1.48   | 4.64  | 0.06  | 0.15  | |
| CXOU J032240.4-371244 | 7 | 46 | 3 22 40.4 | -37 12 44.5 | 3.00 | 31.87  | 9.22  | 1.32  | 3.31  | |
| CXOU J032240.5-371227 | 7 | 16 | 3 22 40.5 | -37 12 26.7 | 3.00 | 32.08  | 10.57 | 1.34  | 3.34  | 01 02 |
| CXOU J032240.7-371151 | 7 | 15 | 3 22 40.7 | -37 11 51.4 | 3.00 | 28.12  | 7.82  | 1.18  | 2.94  | |
| CXOU J032240.8-371224 | 7 | 14 | 3 22 40.8 | -37 12 23.9 | 3.00 | 54.97  | 11.86 | 2.30  | 5.75  | 01 02 |
| CXOU J032241.1-371235 | 7 | 13 | 3 22 41.1 | -37 12 35.4 | 3.00 | 248.57 | 19.24 | 10.39 | 25.97 | 01 07 |
| CXOU J032241.3-371117 | 7 | 45 | 3 22 41.3 | -37 11 17.4 | 3.00 | 28.52  | 7.26  | 1.18  | 2.94  | |
| CXOU J032241.7-371229 | 7 | 11 | 3 22 41.7 | -37 12 29.0 | 3.00 | 805.64 | 31.17 | 33.70 | 84.21 | 01 02 06 |
| CXOU J032241.7-371343 | 7 | 12 | 3 22 41.7 | -37 13 43.2 | 3.00 | 3.95   | 4.47  | 0.16  | 0.40  | |
| CXOU J032241.8-371207 | 7 | 10 | 3 22 41.8 | -37 12 7.4  | 3.00 | 14.45  | 6.61  | 0.60  | 1.51  | |
| CXOU J032241.8-371225 | 7 | 44 | 3 22 41.8 | -37 12 25.1 | 3.00 | 440.46 | 24.47 | 18.40 | 45.98 | 01 06 |
| CXOU J032241.9-371238 | 7 | 42 | 3 22 41.9 | -37 12 38.1 | 3.00 | 52.86  | 13.06 | 2.20  | 5.51  | 01 05 |
| CXOU J032241.9-371305 | 7 | 43 | 3 22 41.9 | -37 13 4.9  | 3.00 | 11.35  | 6.04  | 0.47  | 1.18  | |
| CXOU J032242.0-371218 | 7 | 41 | 3 22 42.0 | -37 12 18.5 | 3.00 | 5.46   | 9.88  | 0.23  | 0.57  | 01 05 |
| CXOU J032242.0-371246 | 7 | 8  | 3 22 42.0 | -37 12 45.5 | 3.00 | 10.00  | 8.96  | 0.41  | 1.03  | 01 |
| CXOU J032242.0-371256 | 7 | 9  | 3 22 42.0 | -37 12 56.1 | 3.00 | 19.77  | 7.13  | 0.82  | 2.04  | 02 |
| CXOU J032242.1-371123 | 7 | 7  | 3 22 42.1 | -37 11 23.3 | 3.00 | 65.08  | 9.61  | 2.69  | 6.72  | |
| CXOU J032242.3-371260 | 7 | 6  | 3 22 42.3 | -37 12 59.5 | 3.00 | 42.73  | 8.44  | 1.77  | 4.43  | 02 |
| CXOU J032242.5-371222 | 7 | 40 | 3 22 42.5 | -37 12 21.5 | 3.00 | 56.18  | 11.76 | 2.34  | 5.84  | 01 02 |
| CXOU J032242.5-371258 | 7 | 5  | 3 22 42.5 | -37 12 57.5 | 3.00 | 18.42  | 6.50  | 0.76  | 1.89  | 02 |
| CXOU J032242.7-371217 | 7 | 4  | 3 22 42.7 | -37 12 17.1 | 3.00 | 6.69   | 9.06  | 0.27  | 0.68  | 01 02 05 |
| CXOU J032242.8-371222 | 7 | 39 | 3 22 42.8 | -37 12 22.2 | 3.00 | 29.86  | 10.43 | 1.23  | 3.09  | 01 02 |
| CXOU J032243.2-371206 | 7 | 38 | 3 22 43.2 | -37 12 6.3  | 3.00 | 3.03   | 4.91  | 0.12  | 0.31  | |
| CXOU J032243.3-371104 | 7 | 3  | 3 22 43.3 | -37 11 3.8  | 3.00 | 36.39  | 7.48  | 1.55  | 3.87  | |
| CXOU J032243.7-371215 | 7 | 75 | 3 22 43.7 | -37 12 15.0 | 3.00 | 25.81  | 7.84  | 1.07  | 2.66  | |
| CXOU J032243.8-371233 | 7 | 2  | 3 22 43.8 | -37 12 32.7 | 3.00 | 21.89  | 7.92  | 0.90  | 2.26  | |
| CXOU J032243.9-371203 | 7 | 37 | 3 22 43.9 | -37 12 2.7  | 3.00 | 8.33   | 5.14  | 0.34  | 0.84  | |
| cxou J032244.2-371053 | 7 | 36 | 3 22 44.2 | -37 10 53.2 | 3.00 | 9.48   | 4.87  | 0.41  | 1.03  | |
| CXOU J032244.8-371210 | 7 | 35 | 3 22 44.8 | -37 12 9.7  | 3.00 | 11.36  | 5.60  | 0.46  | 1.16  | |

(continued)

| source | ccd no. | detect no. | RA (J2000) | DEC (J2000) | source radius (") | net counts | error | counts/ksec | Lx ($10^{38}$ erg s$^{-1}$) | Note |
|---|---|---|---|---|---|---|---|---|---|---|
| CXOU J032245.2-371423 | 7 | 73 | 3 22 45.2 | -37 14 22.7 | 3.00 | 8.78 | 4.29 | 0.52 | 1.30 | 03 |
| CXOU J032245.4-371226 | 7 | 34 | 3 22 45.4 | -37 12 25.5 | 3.00 | 2.00 | 4.17 | 0.08 | 0.20 | |
| CXOU J032245.4-371306 | 7 | 72 | 3 22 45.4 | -37 13 6.2 | 3.00 | 12.63 | 5.24 | 0.51 | 1.29 | |
| CXOU J032245.6-371207 | 7 | 33 | 3 22 45.6 | -37 12 6.7 | 3.00 | 8.11 | 5.01 | 0.33 | 0.83 | |
| CXOU J032245.8-371214 | 7 | 1 | 3 22 45.8 | -37 12 13.8 | 3.00 | 19.58 | 6.21 | 0.80 | 2.00 | |
| CXOU J032247.1-371408 | 7 | 71 | 3 22 47.1 | -37 14 7.5 | 3.00 | 4.25 | 3.62 | 0.23 | 0.57 | 03 |
| CXOU J032247.5-371240 | 7 | 70 | 3 22 47.5 | -37 12 39.7 | 3.00 | 9.46 | 4.73 | 0.38 | 0.95 | |
| CXOU J032248.3-371159 | 7 | 32 | 3 22 48.3 | -37 11 58.9 | 3.00 | 6.68 | 4.30 | 0.27 | 0.68 | |
| CXOU J032249.3-371152 | 7 | 31 | 3 22 49.3 | -37 11 52.0 | 3.00 | 16.62 | 5.46 | 0.68 | 1.71 | |
| CXOU J032249.4-371033 | 7 | 30 | 3 22 49.4 | -37 10 33.0 | 4.10 | 5.23 | 4.75 | 0.21 | 0.51 | |
| CXOU J032250.9-370813 | 7 | 79 | 3 22 50.9 | -37 8 13.3 | 8.18 | 17.98 | 7.54 | 0.78 | 1.95 | |
| CXOU J032251.0-371206 | 7 | 69 | 3 22 51.0 | -37 12 5.8 | 3.21 | 10.68 | 4.86 | 0.43 | 1.08 | |
| CXOU J032251.2-370949 | 7 | 29 | 3 22 51.2 | -37 9 49.2 | 5.43 | 227.86 | 16.54 | 9.46 | 23.65 | 04 |
| CXOU J032252.2-371158 | 7 | 28 | 3 22 52.2 | -37 11 57.9 | 3.68 | 14.45 | 5.47 | 0.59 | 1.47 | |
| CXOU J032253.6-371211 | 7 | 27 | 3 22 53.6 | -37 12 11.1 | 4.03 | 15.28 | 5.58 | 0.62 | 1.56 | |
| CXOU J032254.3-371027 | 7 | 68 | 3 22 54.3 | -37 10 26.8 | 5.54 | 14.82 | 5.91 | 0.65 | 1.62 | |
| CXOU J032259.7-371128 | 7 | 66 | 3 22 59.7 | -37 11 27.7 | 6.48 | 26.05 | 7.35 | 1.09 | 2.72 | |
| CXOU J0323 5.7-371113 | 7 | 65 | 3 23 5.7 | -37 11 13.0 | 9.57 | 92.49 | 11.82 | 3.96 | 9.90 | |

<Within the D25 ellipse of NGC 1317>

| source | ccd no. | detect no. | RA (J2000) | DEC (J2000) | source radius (") | net counts | error | counts/ksec | Lx ($10^{38}$ erg s$^{-1}$) | Note |
|---|---|---|---|---|---|---|---|---|---|---|
| CXOU J032240.7-370518 | 7 | 81 | 3 22 40.7 | -37 5 18.4 | 15.26 | 1.03 | 9.23 | 0.07 | 0.17 | 03 |
| CXOU J032244.3-370614 | 7 | 74 | 3 22 44.3 | -37 6 14.0 | 11.79 | 366.52 | 21.45 | 16.39 | 40.97 | 08 |
| CXOU J032246.1-370552 | 7 | 80 | 3 22 46.1 | -37 5 51.9 | 13.79 | 47.13 | 11.82 | 2.16 | 5.40 | |

<Beyond D25 of NGC 1316 and NGC 1317>

| source | ccd no. | detect no. | RA (J2000) | DEC (J2000) | source radius (") | net counts | error | counts/ksec | Lx ($10^{38}$ erg s$^{-1}$) | Note |
|---|---|---|---|---|---|---|---|---|---|---|
| CXOU J032145.8-370430 | 2 | 3 | 3 21 45.8 | -37 4 29.9 | 40.00 | 113.70 | 19.81 | 4.447 | | |
| CXOU J032147.1-370723 | 2 | 2 | 3 21 47.1 | -37 7 23.4 | 31.57 | 87.04 | 16.67 | 3.177 | | |
| CXOU J032159.4-370515 | 2 | 1 | 3 21 59.4 | -37 5 15.3 | 28.50 | 180.51 | 18.68 | 6.489 | | |
| CXOU J032137.5-371454 | 5 | 1 | 3 21 37.5 | -37 14 53.7 | 34.29 | 384.08 | 32.73 | 17.639 | | |
| CXOU J0322 0.7-371918 | 6 | 16 | 3 22 0.7 | -37 19 18.2 | 22.29 | 197.23 | 17.55 | 8.064 | | |
| CXOU J032211.1-370954 | 6 | 12 | 3 22 11.1 | -37 9 53.8 | 8.56 | 30.20 | 7.66 | 1.111 | | |
| CXOU J032212.6-371212 | 6 | 11 | 3 22 12.6 | -37 12 11.8 | 5.84 | 26.37 | 6.75 | 0.958 | | |
| CXOU J032215.4-371656 | 6 | 7 | 3 22 15.4 | -37 16 55.9 | 8.03 | 78.53 | 10.45 | 3.034 | | |
| CXOU J032225.7-370912 | 7 | 64 | 3 22 25.7 | -37 9 12.0 | 5.22 | 136.60 | 13.09 | 6.268 | | |
| CXOU J032230.8-370736 | 7 | 77 | 3 22 30.8 | -37 7 36.1 | 7.77 | 49.72 | 9.28 | 2.234 | | |
| CXOU J032254.7-370706 | 7 | 67 | 3 22 54.7 | -37 7 6.5 | 11.84 | 74.48 | 11.84 | 3.461 | | |

Note.
01: Sources within 20" from the center of NGC 1316. The source count
    may be contaminated by the diffuse emission.
02: There is another source(s) within the source extraction radius.
03: Sources fall near the edge of the detector. The source count should be considered as a lower limit.
04: a potential ULX source
05: coincident with globular clusters listed in Shaya et al. (1996) and Goudfooij et al. (2001)
06: at the center of NGC 1316
07: close to the center of NGC 1316. Its count may be over-estimated.
08: at the center of NGC 1317
09: a potential super soft X-ray source

Lx (in 0.3-8.0 keV) is determined with D = 18.6 Mpc and ECF = 6.037 (or 9.767) x 10^-15 erg/sec for 1 count/ksec
which is appropriate in BI (or FI) CCDs for a source with photon index = 1.7 and N(H) = 3 x 10^20 cm-2.

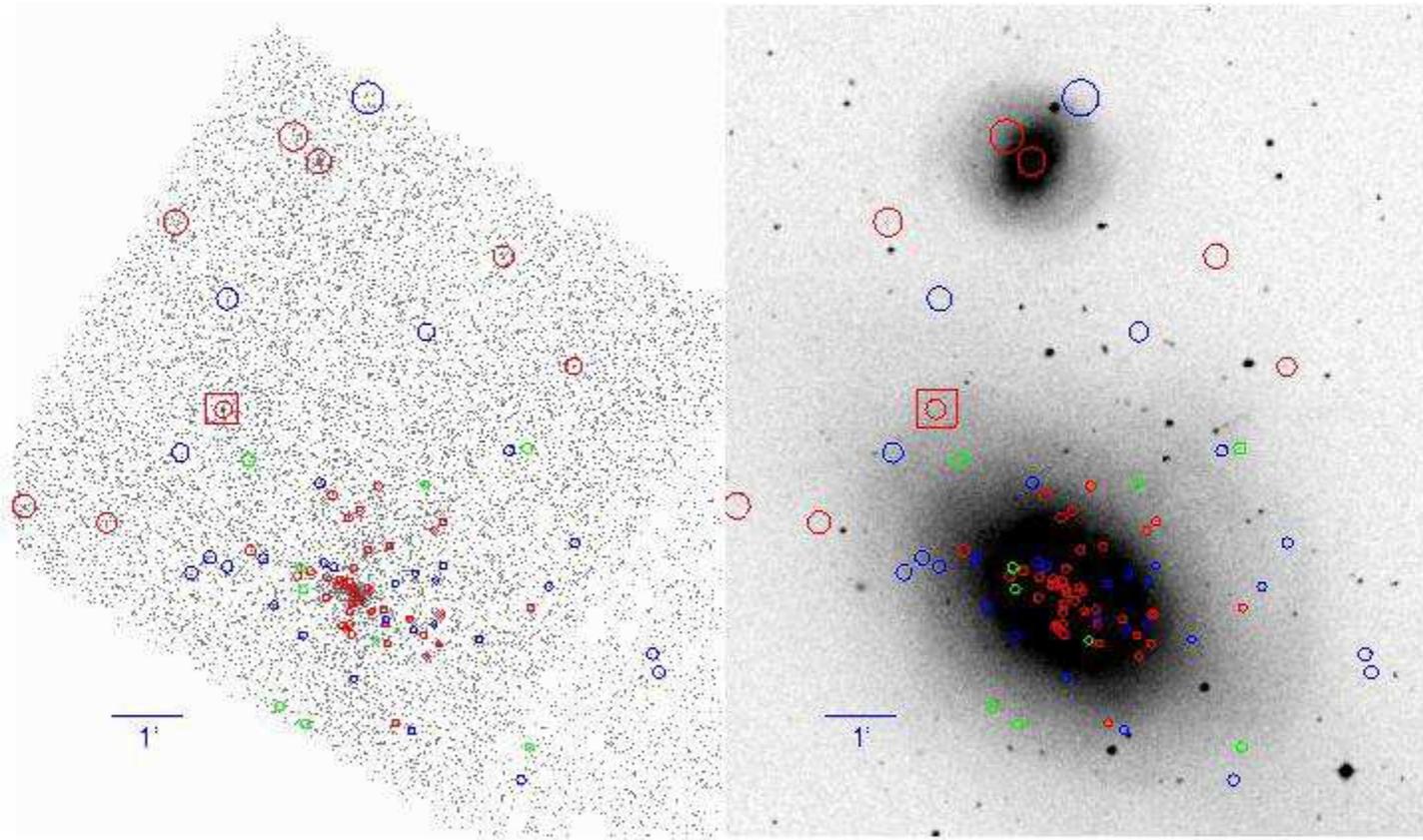

Figure 1. X-ray sources superimposed on the X-ray (left) and on the Digitized Sky Survey optical (right) images. NGC 1316 is at the center of the DSS image and NGC 1317 to the north of NGC 1316. The horizontal bar in the lower left corner indicates 1'. The circle radius indicates the PSF size while the circle color indicates **wavdetect** source significance (red for > 5 , blue for 3-5 and green for < 3). The source marked with a square is a possible ULX source.

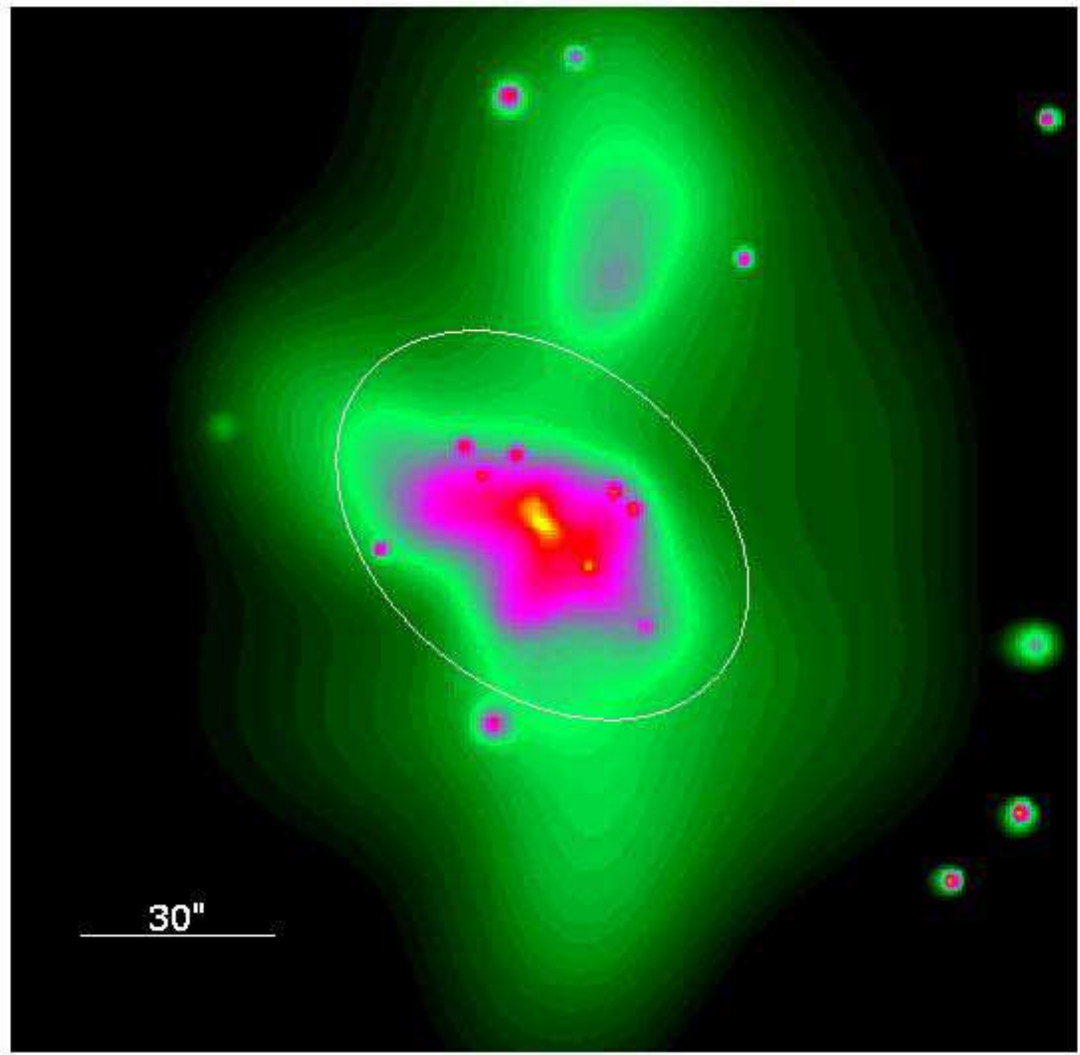

Figure 2. Chandra X-ray image of the central 3'x3' region of NGC 1316. The data were adaptively smoothed. The ellipse indicates the optical figure of NGC 1316 with size equal to 1/10 of $D_{25}$ taken from RC3. The horizontal bar on the lower left corner is 30" long.

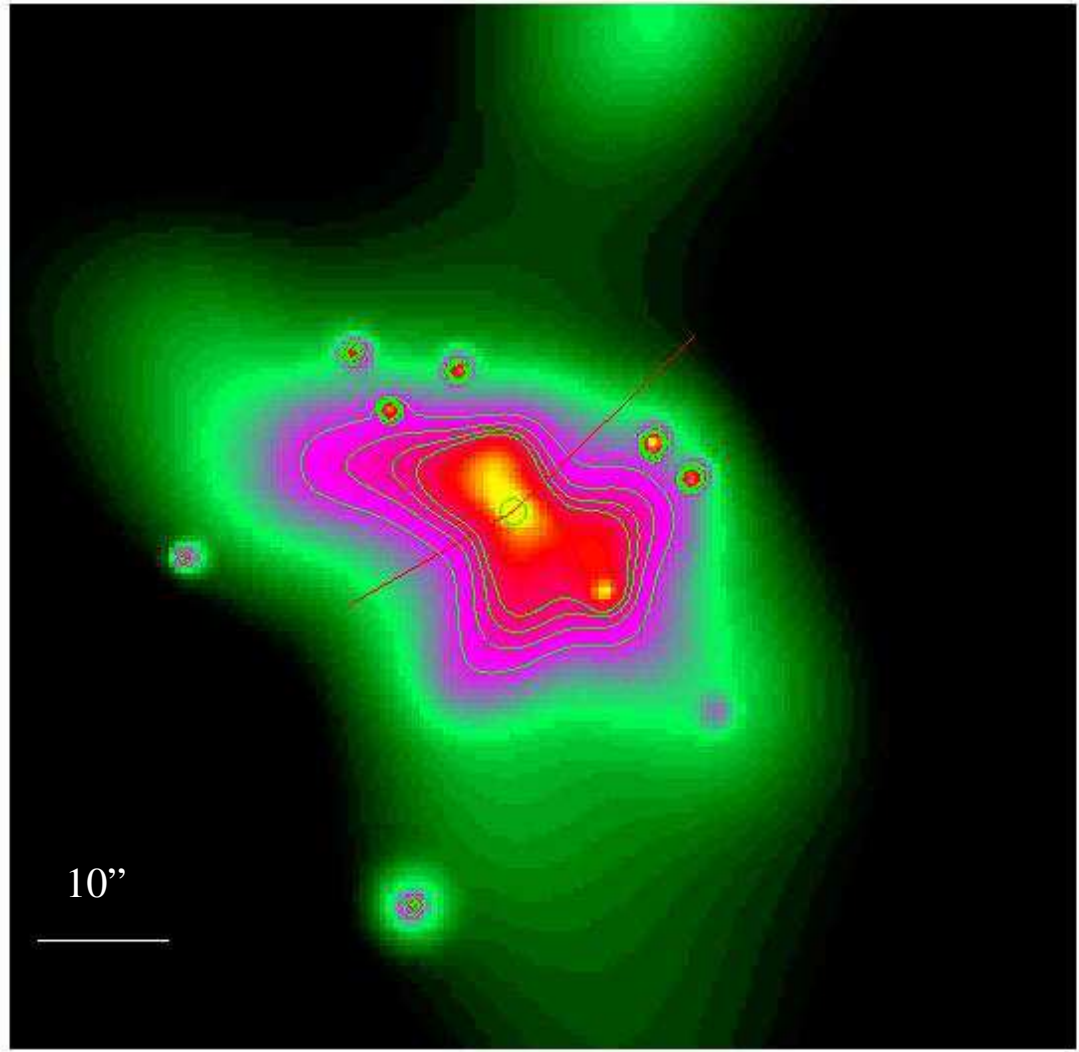

Figure 3. The enlarged view of the central region of NGC 1316 from the Chandra X-ray data (adaptive smoothing applied). The horizontal bar in the lower left corner is 10" long. The location of the nucleus determined in the hard band image is marked by a green circle. A few contours are drawn to illustrate the X-ray valleys along PA~120 and PA~315 where the radio jets (marked as red lines) are propagating.

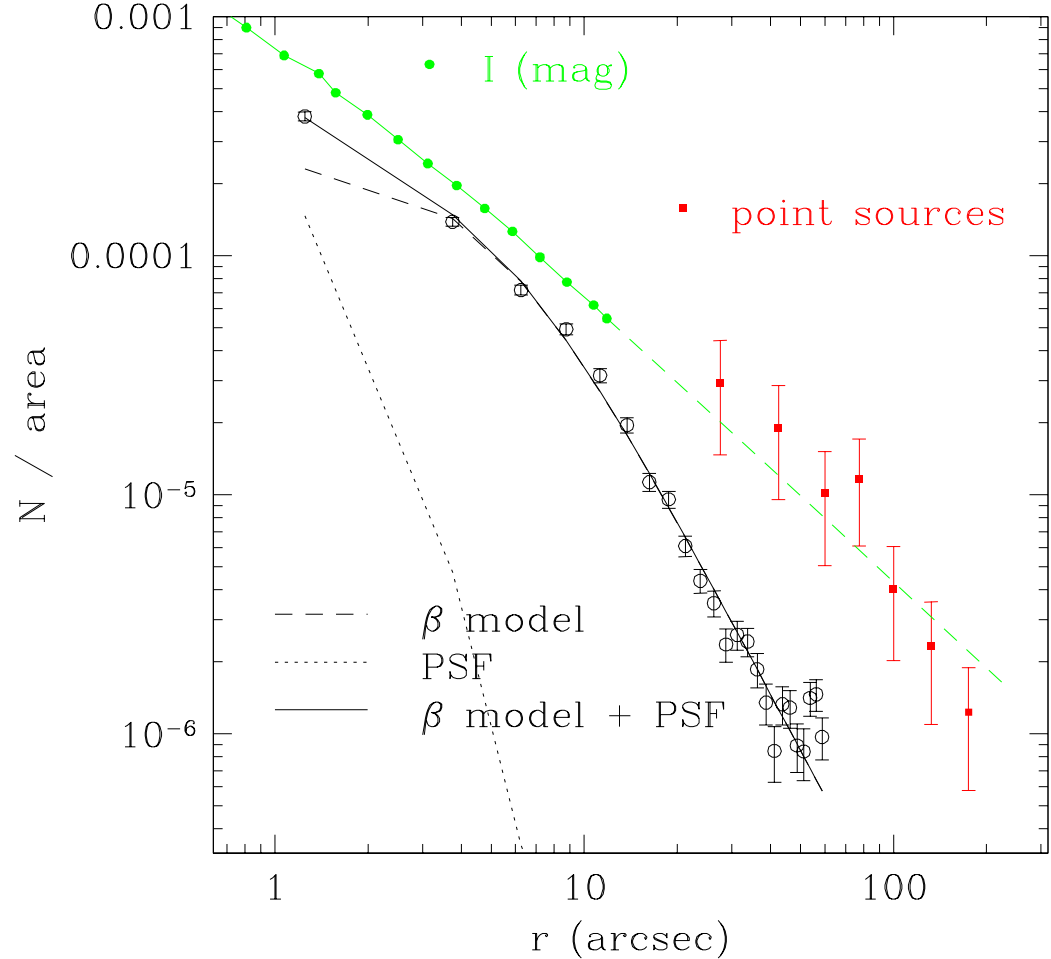

Figure 6. Radial profile of the diffuse X-ray emission (open circles) and point sources (red squares). The solid and dashed lines indicate the beta-model predictions with and without AGN (the dotted line) at the center. For comparison, the I-band radial profile is also plotted in green. The y-axis zero points of the I-band and point source distributions were rescaled to make the two distributions to fit in a single plot. This does not change the radial dependence of the profiles.

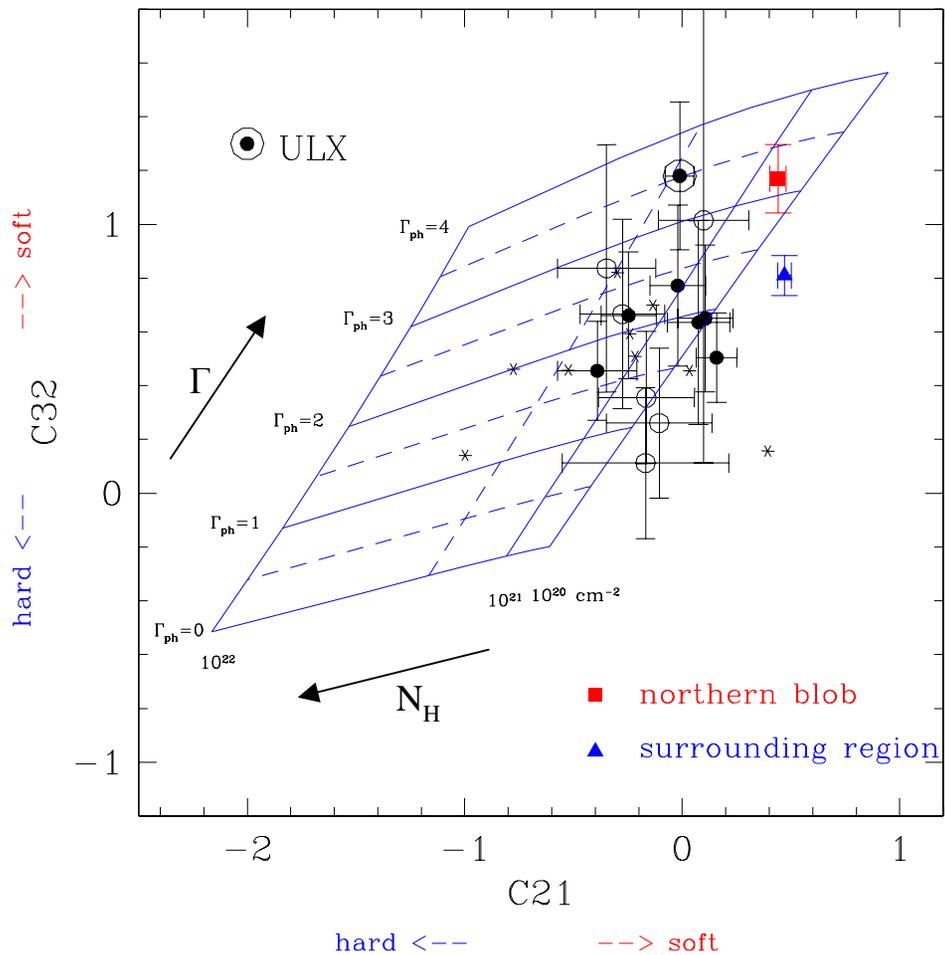

Figure 7. X-ray color-color plot of point sources within D25 of NGC 1316. Sources with more than 50 counts, 30-50 and 20-30 counts are denoted by filled circles, open circles and stars. The brightest source (the ULX) is marked separately. The grid covers photon index $\Gamma_{ph}$ = 0 – 4 (from bottom to top) and N(H) = $10^{20}$ – $10^{22}$ (from right to left). The two points on the right indicate the X-ray colors of the northern blob and the surrounding region at the similar distance from the center.

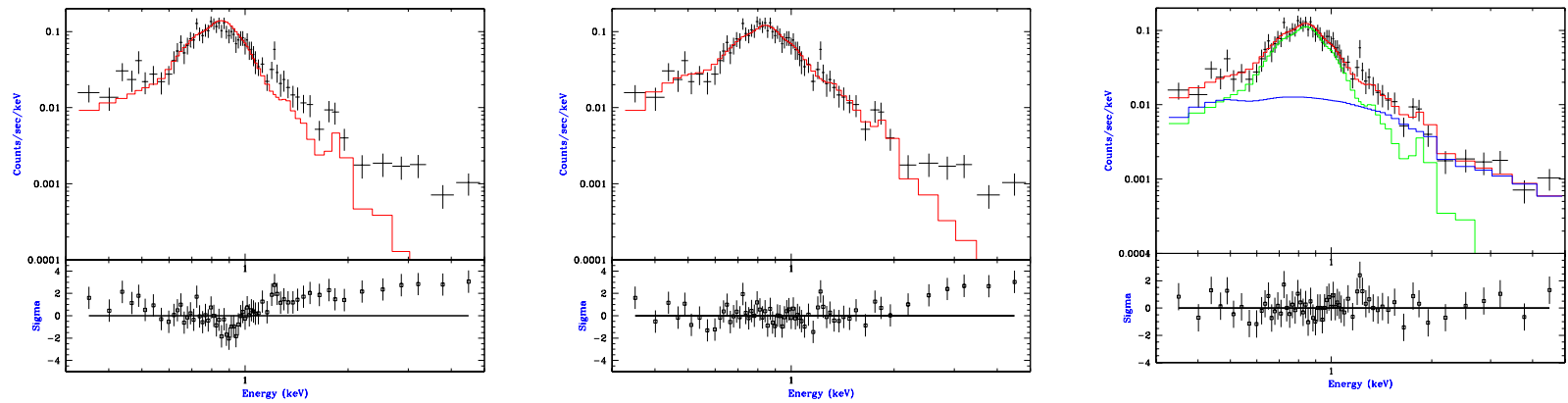

Figure 8. Spectral Fitting at the central bin (r < 5") with (a) a single-component Mekal model with solar abundance, (b) a single-component Mekal model with varying abundance, and (c) a two-component model with Mekal (green) + power law (blue).

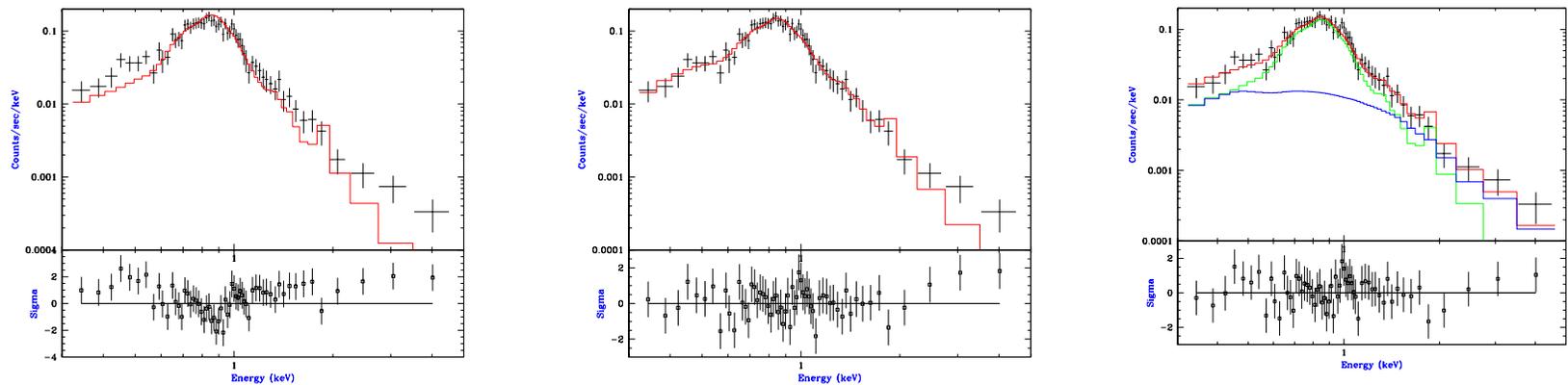

Figure 9. Spectral Fitting at the radial bin of r = 5-15" (all detected point sources excluded from the spectrum) with (a) a single-component Mekal model with solar abundance, (b) a single-component Mekal model with varying abundance, and (c) a two-component model with Mekal (green) + Bremsstrahlung (blue).

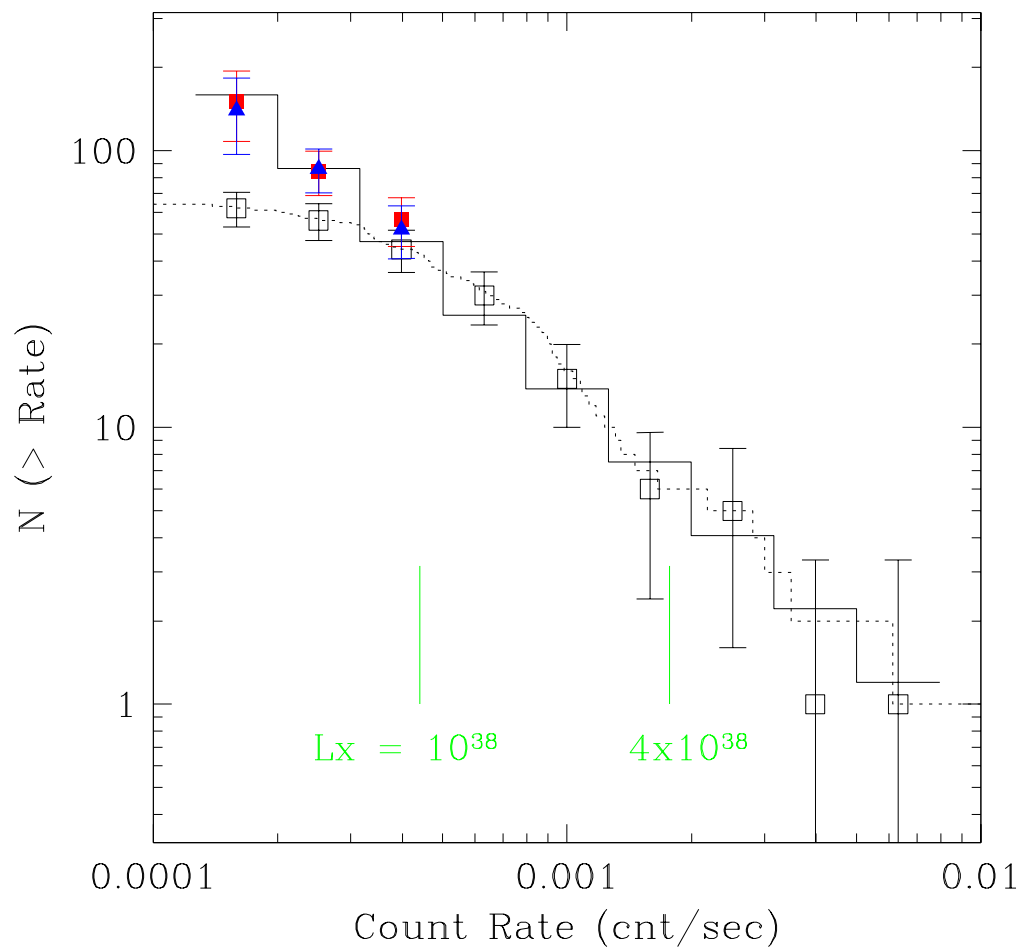

Figure 10. X-ray luminosity function (XLF) of point sources within $D_{25}$ of NGC 1316. The binned and unbinnned data are denoted by open squares and a dotted line, respectively. The solid histogram indicates the model prediction with a single power law, fitted at count rate higher than 0.6 counts/ksec. Three red squares and blue triangles at the lowest Lx bins represent the XLF corrected using two different approaches for estimating incompleteness in source detection (see text in Section 3 and Appendix).

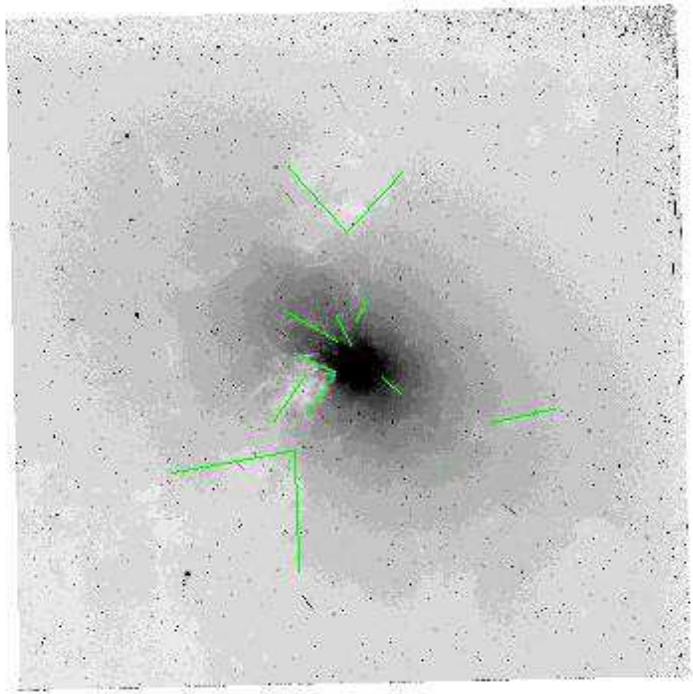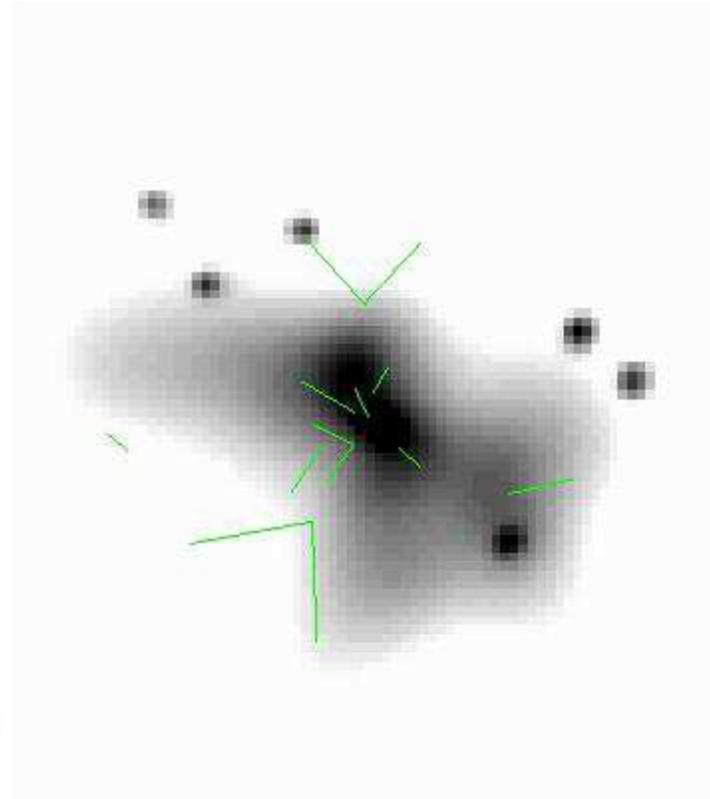

Figure 11. HST WFPC2 image at 4520 A (F450W filter) of the central part (36 x 36") of NGC 1316 (left) and the Chandra X-ray image (right) are plotted in the same scale. The green marks indicate the dust lanes seen in the HST image. North is to the top and east is to the left of the image.

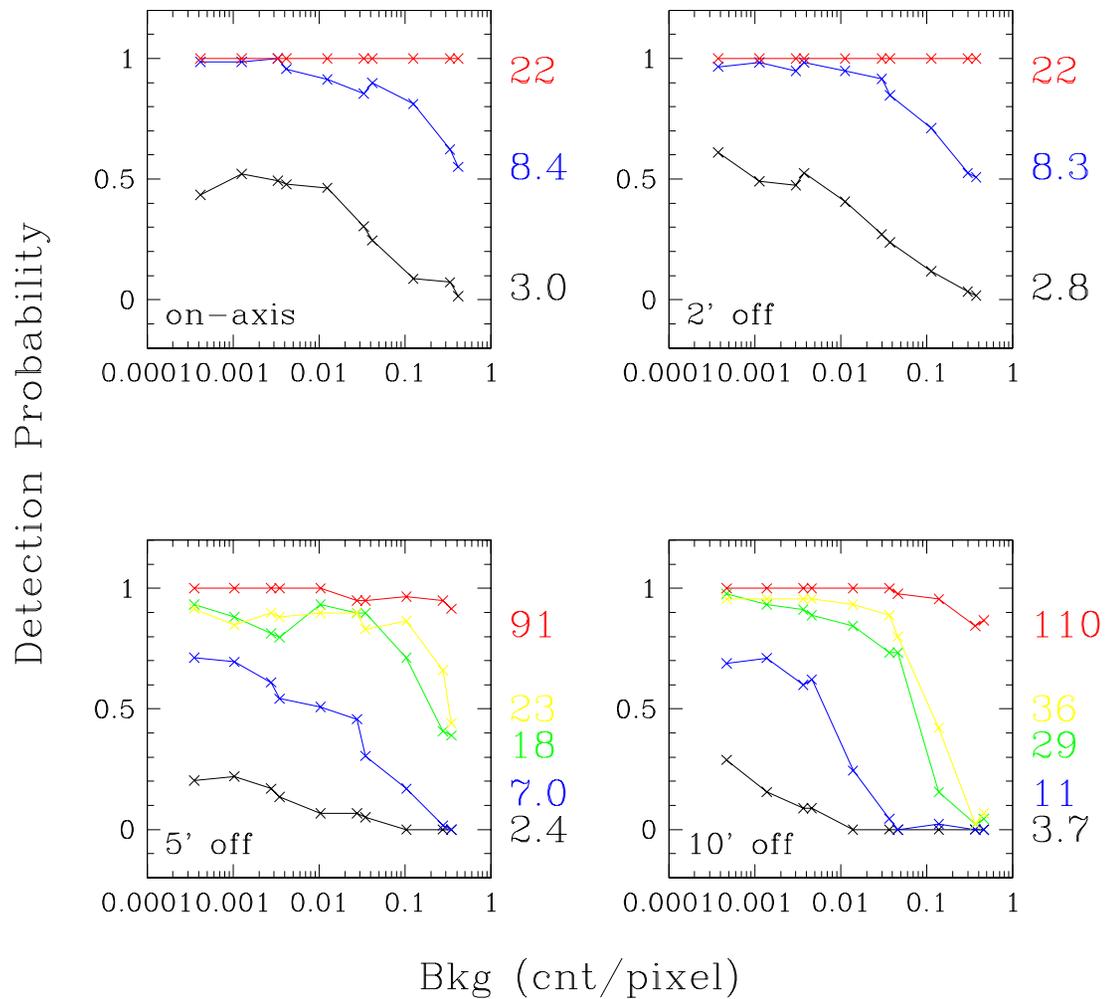

Figure 12. Detection probabilities as a function of background count with various source counts (from a few to ~100) and off-axis distances: (a) on-axis, (b) 2' off-axis, (c) 5' off-axis and (d) 10' off-axis.

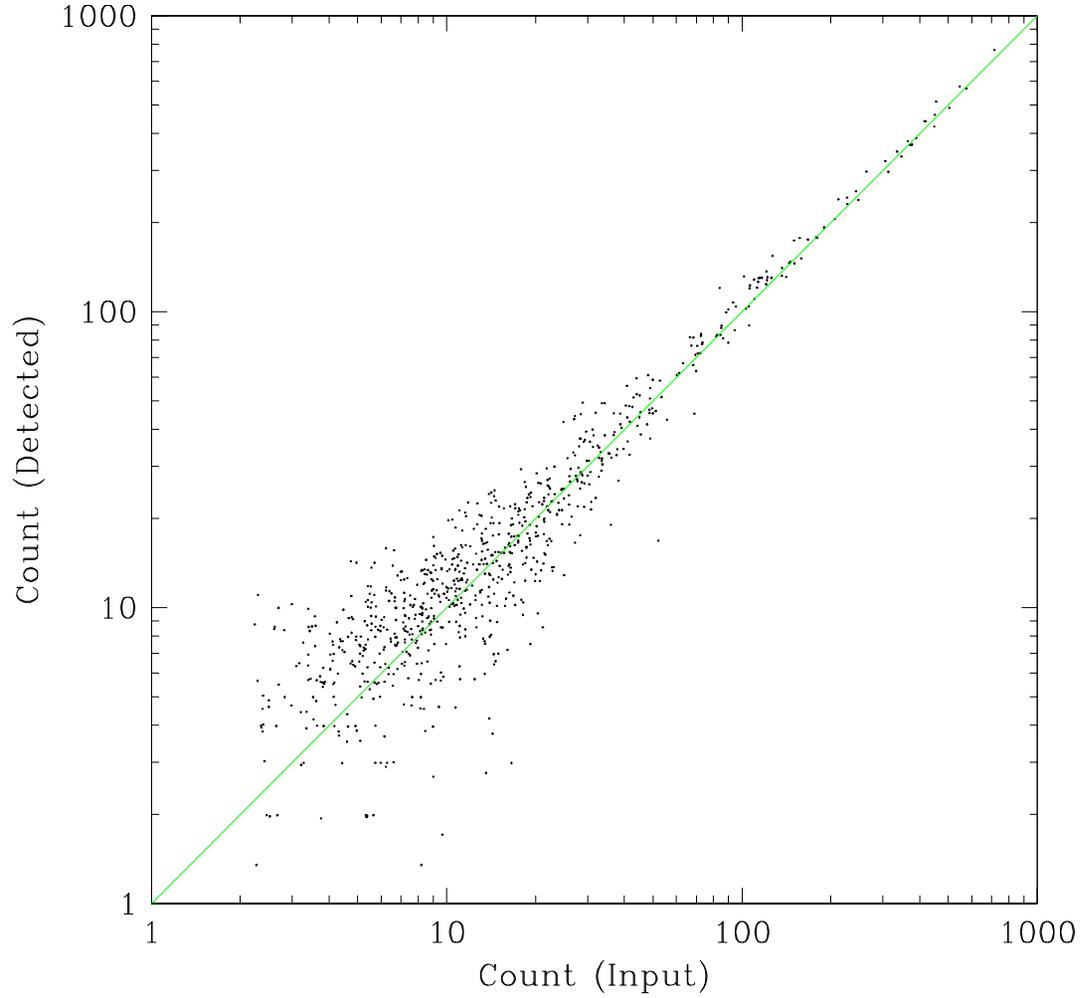

Figure 13. Comparison of detected counts with input counts. The input source flux was selected randomly, according to a single power law, simulated with MARX**,** added to the real image, and then detected by **wavdetect**.

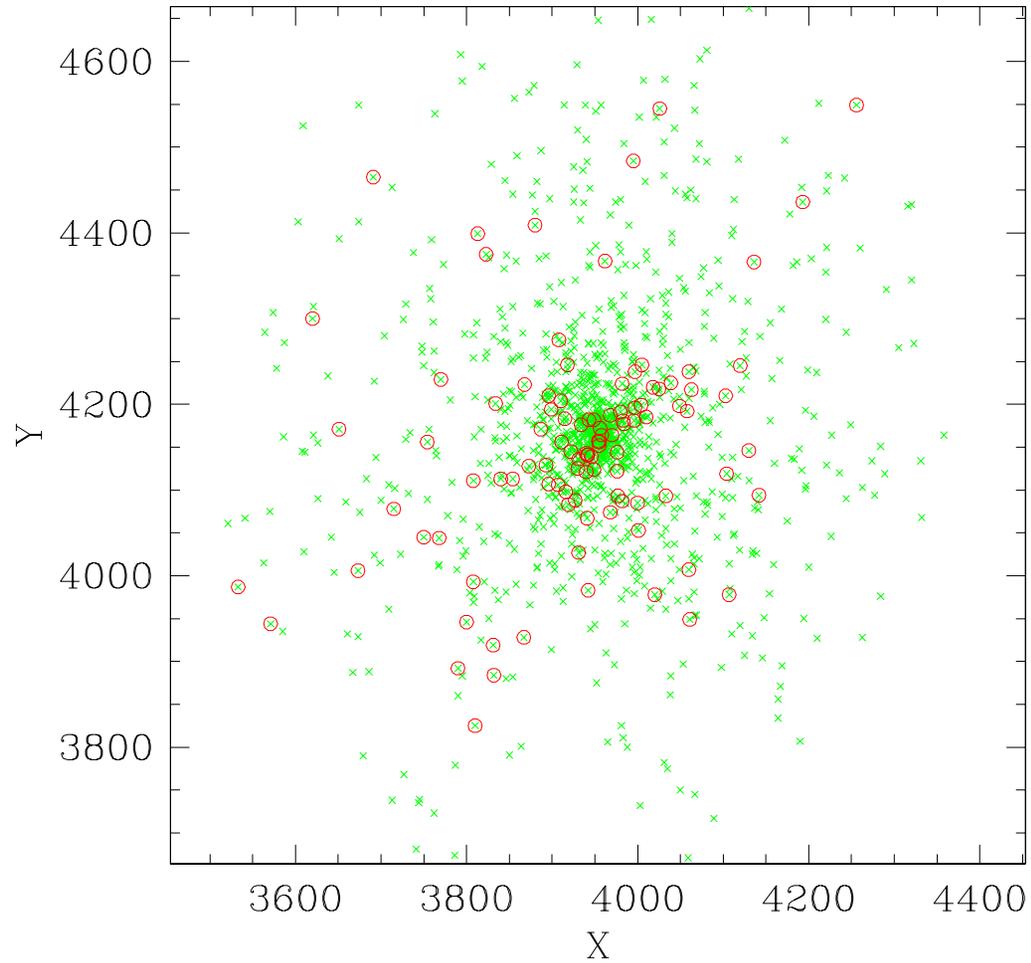

Figure 14. Spatial distribution of simulated sources (green x's). A red circle on a green x indicates a detection. Only 1000 random sources are plotted for visibility. The abscissa and ordinate units are the same as the ACIS sky coordinate, i.e., 1 pixel = 0.492".

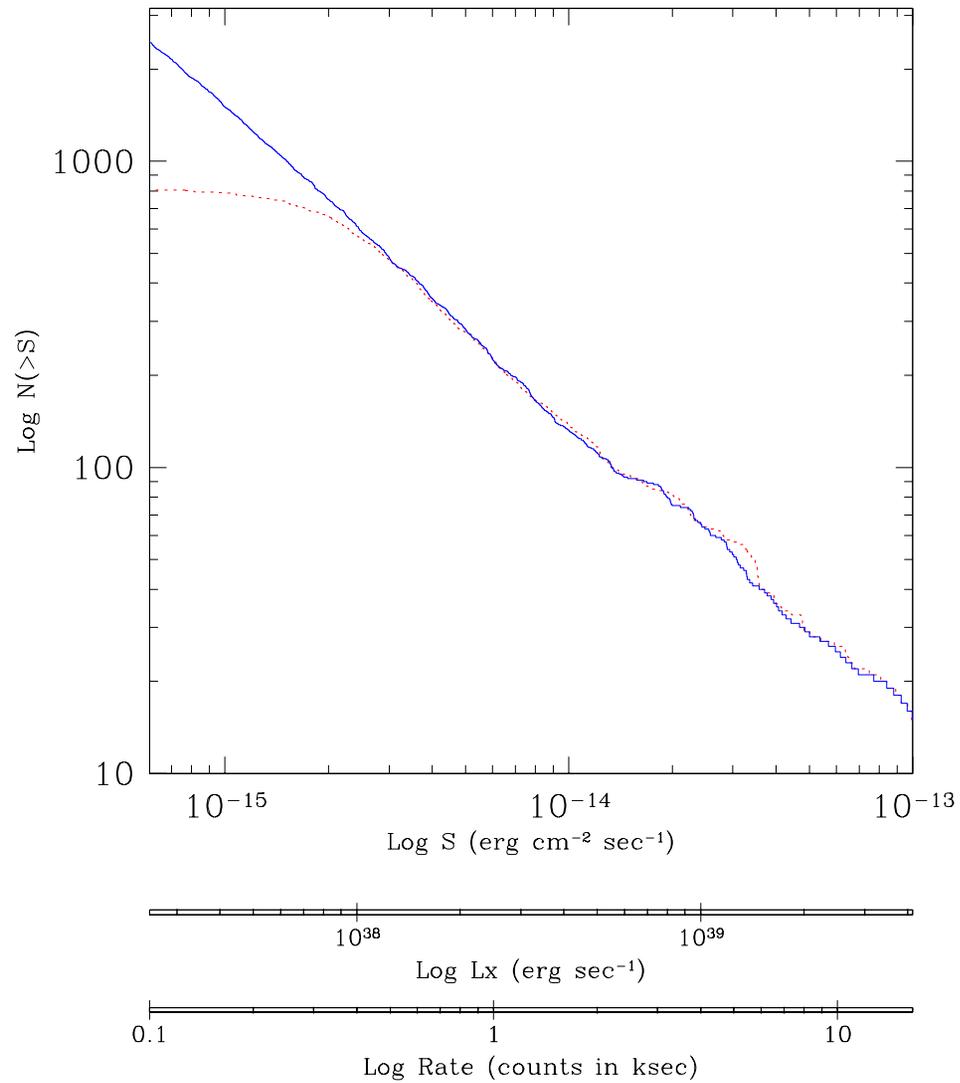

Figure 15. Comparison of input (blue solid line) and measured (red dotted line) XLFs. The ratio of detected to input numbers of sources is applied directly to correct XLF in Figure 10 (a red triangle).